\documentclass[preprint,aps,prd,floatfix,nofootinbib,11pt]{revtex4-2}
\pdfoutput=1
\usepackage{graphicx}
\usepackage{amsfonts}
\usepackage{nicefrac}
\usepackage{xcolor}
\usepackage{natbib}
\usepackage{braket}

\usepackage{hyperref}

\usepackage{amsmath}
\usepackage{rotating}
\usepackage{amssymb,amsthm}
\usepackage{soul}

\usepackage{xcolor}
\usepackage[normalem]{ulem}

\usepackage{latexsym}
\usepackage{graphics}

\usepackage{amsfonts}
\usepackage{amsmath}
\usepackage{rotating}
\usepackage{amssymb}

\usepackage{amsmath}
\usepackage{rotating}
\usepackage{amssymb}
\usepackage{soul}

\usepackage{xcolor}

\usepackage{xcolor}
\def\beq{\begin{eqnarray}}  
	\def\eeq{\end{eqnarray}}

\begin{document}

\title{Static spherically symmetric perfect fluid solutions in teleparallel $F(T)$ gravity}
	
\author{A. Landry}
\email{a.landry@dal.ca}
\affiliation{Department of Mathematics and Statistics, Dalhousie University, Halifax, Nova Scotia, Canada, B3H 3J5}

\begin{abstract}

In this paper, we investigate static spherically symmetric teleparallel $F(T)$ gravity containing a perfect isotropic fluid. We first write the field equations and proceed to find new teleparallel $F(T)$ solutions for perfect isotropic and linear fluids. By using a power-law ansatz for the coframe components, we find several classes of new non-trivial teleparallel $F(T)$ solutions. We also find a new class of teleparallel $F(T)$ solutions for a matter dust fluid. After we solve the field equations for a non-linear perfect fluid. Once again, there are several new exact teleparallel $F(T)$ solutions and also some approximated teleparallel $F(T)$ solutions. All these classes of new solutions may be relevant for future cosmological and astrophysical applications.

\end{abstract}

\maketitle

\newpage

\tableofcontents


\section{Introduction}

There are a number of alternative theories of gravity to General Relativity (GR). The $F(T)$-type teleparallel theories of gravity are very promising \cite{Aldrovandi_Pereira2013,Bahamonde:2021gfp,Krssak:2018ywd}. In this theory, the geometry is characterized by the torsion which is a function of the coframe, ${\bf h}^a$, derivatives of the coframe, and a zero curvature and metric compatible spin-connection one-form $\omega^a_{~b}$.  Hence in teleparallel gravity it is necessary to work with a frame basis instead of a metric tensor. In such theories the role of symmetry is no longer as clearly defined as in pseudo-Riemannian geometry, where symmetry is defined in terms of an isometry of the metric or Killing Vectors (KVs). In GR, the Riemannian geometry is completely defined by the curvature of a Levi-Civita connection and calculated from the metric. But it is not really the case for some alternative theories, in particular for teleparallel $F(T)$-type gravity.

The development of a frame based approach for determining the symmetries of a spacetime has been explored  \cite{chinea1988symmetries, estabrook1996moving, papadopoulos2012locally}. A possible complication arises due to the possible existence of a non-trivial linear isotropy group: a Lie group of Lorentz frame transformations keeping the associated tensors of the geometry invariant. If a given spacetime has a non-trivial linear isotropy group, determining the group of symmetries requires solving a set of inhomogeneous differential equations \cite{olver1995equivalence}:
\begin{equation}
\mathcal{L}_{{\bf X}} {\bf h}^a = \lambda^a_{~b} {\bf h}^b \text{ and } \mathcal{L}_{{\bf X}} {\omega}^a_{~bc} = 0, \label{Intro2}
\end{equation}
where ${\bf h}^a$ is the orthonormal coframe basis, $\lambda^a_{~b}$ is a Lie algebra generator of Lorentz transformations and ${\omega}^a_{~bc}$ are the components of the spin-connection.

In ref. \cite{MCH}, Coley et al. introduced a new approach to determine the symmetries of any geometry based on an independent frame and connection which admits the torsion tensor and the curvature tensor as geometric objects. In these theories, the connection is an independent object. They call any geometry where the non-metricity and curvature tensors vanish a {\it teleparallel geometry}. The approach relies on the existence of a particular class of invariantly defined frames known as symmetry frames, which facilitates the solving the differential equations arising from eqn \eqref{Intro2}, by fixing the $\lambda^a_{~b}$ in an invariant way.

This assumes an orthonormal frame where the gauge metric is $g_{ab} = diag[-1,1,1,1]$. The spin-connection $\omega^a_{~bc}$, is defined in terms of an arbitrary Lorentz transformation, $\Lambda^a_{~b}$ through the equation
\begin{equation}
\omega^a_{~bc} = \Lambda^a_{~d} {\bf h}_c((\Lambda^{-1})^d_{~b}). \label{TP:con}
\end{equation}
A particular subclass of teleparallel gravitational theories is dynamically equivalent to GR and is called the Teleparallel Equivalent to General Relativity (TEGR), which is based on a torsion scalar $T$ constructed from the torsion tensor \cite{Aldrovandi_Pereira2013}. The most common generalization of TEGR is $F(T)$-type teleparallel gravity, where $F$ is an arbitrary function of the torsion scalar $T$ \cite{Ferraro:2006jd, Ferraro:2008ey, Linder:2010py}. In the {\it covariant} approach to $F(T)$-type gravity, the teleparallel geometry is defined in a gauge invariant manner as a geometry where the spin-connection has zero curvature and zero non-metricity. The spin-connection will vanish in the special class of proper frames, and will be non-zero in all other frames \cite{Lucas_Obukhov_Pereira2009,Aldrovandi_Pereira2013,Krssak:2018ywd}. Therefore, the resulting teleparallel gravity theory has Lorentz covariant FEs and is locally Lorentz invariant \cite{Krssak_Pereira2015}. A proper frame is not invariantly defined since it is defined in terms of the connection, which is not a tensorial quantity, which leads to  a number of problems when using such a frame to determine symmetries.

There are several papers in the literature about static and non-static spherically symmetric solutions in teleparallel $F(T)$ gravity \cite{golov1,golov2,golov3,debenedictis,SSpaper,TdSpaper,baha1,bahagolov1,awad1,baha6,nashed5,pfeifer2,elhanafy1,benedictis3,baha10,baha4}. There are several perturbative solutions in TEGR (Teleparallel Equivalent of General Relativity) and there are some power-law $F(T)$ solutions with power-law frame components (see \cite{golov1,golov2,golov3,debenedictis} and references within). These papers essentially use the Weitzenback gauge (leading to proper frames) because the antisymmetric FEs are trivially satisfied, but there are arising some extra degrees of freedom (DoF) by imposing the zero spin-connection. This requirement leads to only symmetric parts of FEs and the presented solutions are essentially limited to power-law in $F(T)$ and frame components by using a complex coframe. Beyond these considerations, even if the symmetric parts of FEs and its solutions are similar between the different gauges, the fact remains that the extra DoF potential issue associated with the proper frame should be resolved by a frame changing. For this requirement, it is necessary to go towards a frame where the spin-connections can be found by solving the non-trivial antisymmetric parts of FEs. From there, all the DoFs will be covered by all the FEs and the solutions will be found by a non-trivial approach for spin-connection and coframe components and then for $F(T)$ solutions.

For rectifying this extra DoF potential issue and going further than power-law $F(T)$ solutions, there is a paper on general Teleparallel spherically symmetric geometries with an emphasis on vacuum solutions and possible additional symmetry structures \cite{SSpaper}. They found the general FEs in an orthonormal gauge assuming a diagonal frame and a non-trivial spin-connection, leading to specific antisymmetric parts of FEs and then to well-determined symmetric parts of FEs without extra DoF. There are some specific symmetry structures such as static (radial coordinate dependent), Kantowski-Sacks (KS) (time coordinate dependent) and an additional affine symmetry called $X_4$. For static geometries, the study is restricted to find the $F(T)$ solutions in the vacuum. They found more power-laws solutions, but also more general $F(T)$ solutions such as products, quotient, exponential and/or a mix of these type of functions. In this case, the $X_4$ symmetry will be defined by the time-coordinate derivative $\partial_t$ leading to radial coordinate dependence for coframes, spin-connections and FEs.

For non-vacuum spherically symmetric $F(T)$ solutions, there are in principle several possible types of energy-momentum sources. The most interesting are the perfect isotropic cosmological and astrophysical fluids and there are some teleparallel $F(T)$ solutions such as Bahamonde-Camci's \cite{baha1}. In this paper, there are some specific power-law $F(T)$ solutions leading to some specific types of fluid where they find specific expressions for $P$ and $\rho$. But this type of approach is restrictive because this supposes first a power-law $F(T)$ solution and then they look for possible $\rho$ and $P$. Alternatively for finding new solutions, a different approach would assume an energy-momentum source with an equation of state (EoS) (relation between $P$ and $\rho$ as $P=P(\rho)$) and then find all possible $F(T)$ solutions satisfying the FEs with these EoS relations.

For this paper, we assume a static ($r$-coordinate dependent) spherically symmetric teleparallel geometry in an orthonormal gauge as defined in ref \cite{SSpaper}. We will focus on finding non-vacuum static spherically symmetric teleparallel $F(T)$ solutions. After a brief summary of the static spherically symmetric teleparallel geometry and FEs in section \ref{sect2}, we will find in section \ref{sect3} several possible $F(T)$ solutions for the linear and isotropic perfect fluids. In section \ref{sect33}, we will do the same with a dust fluid, because this special case arises from the conservation laws. In section \ref{sect4}, we will solve FEs and find some $F(T)$ solutions for a non-linear perfect fluid.

We will use the notation as follows: the coordinate indices are $\mu, \nu, \ldots$ and the tangent space indices are $a,b,\ldots$ as in ref \cite{MCH}. The spacetime coordinates will be $x^\mu$. The frame fields are denoted as ${\bf h}_a$ and the dual coframe one-forms are ${\bf h}^a$. The vierbein components are $h_a^{~\mu}$ or $h^a_{~\mu}$. The spacetime metric is $g_{\mu \nu}$ and the Minkowski tangent space metric is $\eta_{ab}$. For a local Lorentz transformation leaving $\eta_{ab}$ unchanged, we write $\Lambda_a^{~b}(x^\mu)$. The spin-connection one-form $\omega^a_{~b}$ is defined as $\omega^a_{~b} = \omega^a_{~bc} {\bf h}^c$. The curvature and torsion tensors are, respectively, $R^a_{~bcd}$ and $T^a_{~bc}$. Covariant derivatives with respect to a metric-compatible connection is denoted using a semi-colon, $T_{abc;e}$. 


\section{Summary of teleparallel spherically symmetric spacetimes and field equations}\label{sect2}

\subsection{Teleparallel static spherical symmetry}

The teleparallel spherically symmetric spacetimes were defined and discussed in detail with all necessary justifications in ref \cite{SSpaper}. From this last paper, there is a new coordinate system to ''diagonalize'' the frame described by a static spherically symmetric vierbein satisfying eqn \eqref{Intro2} is:
\begin{equation}
h^a_{~\mu} = \text{Diag}\left[ A_1(r),\, A_2(r),\,A_3(r),\,A_3(r) \sin(\theta)\right]. \label{VB:SS}
\end{equation}
This frame choice in eqn \eqref{VB:SS} is an invariant symmetry frame, and the most general static spherically symmetric spin-connection is \cite{SSpaper}:
\beq \begin{aligned} & \omega_{341} = W_1(r),~ \omega_{342} = W_2(r), ~\omega_{233} = \omega_{244} = W_3(r),~ \omega_{234} = -\omega_{243} = W_4(r), ~ \omega_{121} = W_5(r) \\
& \omega_{122} = W_6(r),~ \omega_{133} = \omega_{144} = W_7(r),~\omega_{134} = -\omega_{143} = W_8(r), \omega_{344} = - \frac{\cos(\theta)}{A_3 \sin(\theta)}. \end{aligned} \label{Con:SS} \eeq
We determine the most general connection by imposing the flatness condition on the geometry. The resulting eqns can be solved so that any spherically symmetric teleparallel geometry is defined by the three arbitrary functions in the vierbein described by eqn \eqref{VB:SS} and by the following spin-connection components \cite{SSpaper}:
\beq \begin{aligned} 
W_1 &= 0,~ W_2 = -\frac{\chi'}{A_2},~ W_3 &= \frac{\cosh(\psi)\cos(\chi)}{A_3}, W_4 = \frac{\cosh(\psi)\sin(\chi)}{A_3},\\
W_5 &= 0, W_6 = -\frac{\psi'}{A_2},~ W_7 &= \frac{\sinh(\psi) \cos(\chi)}{A_3}, W_8 = \frac{\sinh(\psi) \sin(\chi)}{A_3},  
\end{aligned} \label{SS:TPcon} \eeq

\noindent where $\chi(r)$ and $\psi(r)$ are arbitrary functions of the radial coordinate $r$ and then $\chi'=\partial_r\chi(r)$ and $\psi'=\partial_r\psi(r)$.  We will subsequently find via the antisymmetric parts of the FEs the exact expressions for $\chi(r)$ and $\psi(r)$, and therefore the eqns \eqref{Con:SS} and \eqref{SS:TPcon} components.

Any choice of the arbitrary functions, $\psi$ and $\chi$, picks out a unique teleparallel geometry, as any change in the form of the spin-connection which could affect the form of $\psi$ or $\chi$ leads to a change in the form of the vierbein. For a given pair of functions, the invariantly defined frame up to the linear isotropy group $\bar{H}_q$ arising from the Cartan-Karlhede (CK) algorithm could be computed to provide further sub-classification. We note that there are only five arbitrary functions required to specify a geometry: $A_1, A_2, A_3, \psi$ and $\chi$ \cite{SSpaper}. We note that special subclasses of these teleparallel geometries have been studied earlier in teleparallel gravity using the Killing equations for an arbitrary spherically symmetric metric and using the non-invariant proper frame approach \cite{pfeifer2,sharif2009teleparallel,hohmann2019modified,pfeifer2022quick}.

\subsection{Summary of teleparallel field equations}

The teleparallel action integral is \cite{Aldrovandi_Pereira2013,Bahamonde:2021gfp,Krssak:2018ywd,SSpaper}:
\begin{equation}\label{1000}
S_{F(T)} = \int\,d^4\,x\,\left[\frac{h}{2\kappa}\,F(T)+\mathcal{L}_{Matter}\right].
\end{equation}
By applying the least-action principle to the eqn \eqref{1000}, we obtain the symmetric and antisymmetric parts of FEs \cite{SSpaper}:
\begin{subequations}
\begin{eqnarray}
\kappa\,\Theta_{\left(ab\right)} &=& F'\left(T\right) \overset{\ \circ}{G}_{ab}+F''\left(T\right)\,S_{\left(ab\right)}^{\;\;\;\mu}\,\partial_{\mu} T+\frac{g_{ab}}{2}\,\left[F\left(T\right)-T\,F'\left(T\right)\right],  \label{1001a}
\\
0 &=& F''\left(T\right)\,S_{\left[ab\right]}^{\;\;\;\mu}\,\partial_{\mu} T, \label{1001b}
\end{eqnarray}
\end{subequations}
where $\overset{\ \circ}{G}_{ab}$ is the Einstein tensor, $\Theta_{\left(ab\right)}$ the energy-momentum, $T$ the Torsion scalar, $g_{ab}$ the gauge metric, $S_{ab}^{\;\;\;\mu}$ the Superpotential (Torsion dependent) and $\kappa$ the coupling constant. The canonical Energy-Momentum is obtained from $\mathcal{L}_{Matter}$ term of eqn \eqref{1000} and defined as:
\begin{align}\label{1001ca}
\Theta_a^{\;\;\mu}=\frac{1}{h} \frac{\mathcal{L}_{Matter}}{\delta h^a_{\;\;\mu}}.
\end{align}
The antisymmetric and symmetric parts of eqn \eqref{1001ca} are respectively \cite{SSpaper}:
\begin{equation}\label{1001c}
\Theta_{[ab]}=0,\qquad \Theta_{(ab)}= T_{ab},
\end{equation}
where $T_{ab}$ is the symmetric part of energy-momentum tensor. In this eqn \eqref{1001c}, we see that $\Theta_{ab}$ is a symmetric physical quantity. This eqn \eqref{1001c} is valid especially for the case where the matter field interacts with the metric $g_{\mu\nu}$ associated with the coframe $h^a_{\;\;\mu}$ and the gauge $g_{ab}$, and is not intricately coupled to the $F(T)$ gravity. This consideration is valid in the case of this paper, because there is an absence of hypermomentum (i.e. $\mathfrak{T}^{\mu\nu}=0$ as defined in ref \cite{golov3}.). The conservation of energy-momentum in teleparallel gravity states that $\Theta_a^{\;\;\mu}$ must satisfy the following relation as \cite{Aldrovandi_Pereira2013,Bahamonde:2021gfp}:
\begin{align}\label{1001e}
\nabla_{\nu}\left(\Theta^{\mu\nu}\right)=0,
\end{align}
where $\nabla_{\nu}$ is th covariant derivative and $\Theta^{\mu\nu}$ is the conserved energy-momentum tensor. This eqn \eqref{1001e} is the same conservation of energy-momentum expression as GR. Satisfying the eqn \eqref{1001e} is automatically required by the previous equations in cases of null hypermomentum ($\mathfrak{T}^{\mu\nu}=0$ case). In non-zero hypermomentum situations ($\mathfrak{T}^{\mu\nu}\neq 0$ case), we will need to satisfy more complex conservation equations than eqn \eqref{1001e} as showed in ref. \cite{golov3}.

For a perfect and isotropic fluid with any EoS (linear or not), the $T_{ab}$ tensor is defined as \cite{hawkingellis1,coleybook}:
\begin{align}\label{1001d}
T_{ab}= \left(P(\rho(r))+\rho(r)\right)\,u_a\,u_b+g_{ab}\,P(\rho(r)),
\end{align}
where $P(\rho(r))$ is the EoS in terms of the static fluid density $\rho(r)$ and $u_a=(1,\,0,\,0,\,0)$ for a stationary fluid. In some astrophysical applications, the pressure is sometimes modeled with a radial and a tangential components, especially for stellar modeling \cite{hawkingellis1,coleybook}. But this is not the specific purpose of this paper.


\subsection{General static spherically symmetric perfect fluid field equations}

For static spherically symmetric case, the antisymmetric FEs in terms of eqns \eqref{Con:SS} and \eqref{SS:TPcon} components  are \cite{SSpaper}: 
\begin{eqnarray}\label{1004}
0 &=& \frac{F''\left(T\right)\,\partial_r\,T}{\kappa\,A_2\,A_3}\,\left[\cos\,\chi\,\sinh\,\psi\right] \quad \text{and}\quad 0 = \frac{F''\left(T\right)\,\partial_r\,T}{\kappa\,A_2\,A_3}\,\left[\sin\,\chi\,\cosh\,\psi\right] .
\end{eqnarray}
Assuming $T \neq$ constant, eqn. \eqref{1004} admits only the solution $\sin\,\chi=0$ and $\sinh\psi=0$, with $\chi = n\,\pi$ and $\psi=0$ where $n \in \mathbb{Z}$ is an integer and $\cos\,\chi=\cos\left(n\,\pi\right)=\pm 1 = \delta$. Substituting these expressions for $\chi$ and $\psi$ into eqn \eqref{SS:TPcon}, we obtain that $W_3(r)=\frac{\delta}{A_3}$ is the only non-zero component (i.e. $W_i=0$ for all $i \neq 3$.) and find as eqn \eqref{Con:SS} for the non-vanishing spin-connection components:
\begin{align}\label{1050}
{\omega_{233} = \omega_{244} = \frac{\delta}{A_3},~ \omega_{344} = - \frac{\cos(\theta)}{A_3 \sin(\theta)}. }
\end{align}
The eqn \eqref{1050} for non-vanishing spin-connections components goes in the same direction and improves the expressions obtained recently in refs \cite{Krssak:2018ywd,golov3}.

\noindent By using the solution for eqn \eqref{1004}, we find the three equations for the symmetric FEs \cite{SSpaper}:
\begin{subequations}
\begin{align}
\partial_r\,\left[\ln\,F'(T(r))\right]=& \frac{g_1(r)}{k_1(r)} , \label{2201a}
\\
\kappa\,\left[\rho+P\right] =& -2\,F''(T)\,\left(\partial_r\,T\right)\,k_2(r)+2\,F'(T)\,g_2(r) , \label{2201b}
\\
\kappa\,\rho =& -\frac{F(T)}{2} -2\,F''(T)\,\left(\partial_r\,T\right)\,k_3(r)+2\,F'(T)\,g_3(r) , \label{2201c}
\end{align}
\end{subequations} 
where $g_i$ and $k_i$ components are expressed in appendix \ref{appena}. In addition, because we are not in vacuum, we need to satisfy the static conservation law \cite{SSpaper}
\begin{align}
&\left[\rho+P\right]\,\partial_r\left(\ln\,A_1\right)+\partial_r\,P=0, \label{2201d}
\end{align}
where $A_1'=\partial_r\,A_1(r)$ and $P'=\partial_r\,P(r)$ are $r$ radial coordinate derivatives. For $P=-\rho$, we obtain from eqn \eqref{2201d} that $P=P_0=-\rho_0=$ constant. In this case, we will obtain with eqn \eqref{2201a} to \eqref{2201c} the vacuum solutions, but with a $-2\kappa\,\rho_0$ shifting inside the $F(T)$ solutions \cite{SSpaper,TdSpaper}. From FEs components in appendix \ref{appena}, \textbf{we have $k_2(r)=k_3(r)$ for all $A_i$, $i=1,\,2,\,3$}. By also substituting eqn \eqref{2201a} into eqns \eqref{2201b} and \eqref{2201c}, we find that the FEs become:
\begin{subequations}
\begin{align}
F'(T(r))=& F'(T(0))\,\exp\left[\int\,dr'\,\frac{g_1(r')}{k_1(r')}\right] , \label{2202a}
\\
\kappa\,\left[\rho+P\right] =& 2\,F'(T)\,\left[-\left(\frac{g_1(r)}{k_1(r)}\right)\,k_2(r)+g_2(r)\right] , \label{2202b}
\\
\kappa\,\rho =& -\frac{F(T)}{2} + 2\,F'(T)\,\left[-\left(\frac{g_1(r)}{k_1(r)}\right)\,k_2(r)+g_3(r)\right] . \label{2202c}
\end{align}
\end{subequations} 
From there, we have to solve the eqns \eqref{2201d} to \eqref{2202c} for non-vacuum solutions by using the torsion scalar expression and the $g_i$ and $k_i$.


\section{Perfect linear fluid solutions}\label{sect3}

As the first case of a non-vacuum solution with an isotropic fluid having a linear EoS, we have $P(r)=\alpha\,\rho(r)$ with $-1 < \alpha <0$ and $0<\alpha \leq 1$ (i.e. $\alpha \neq 0$), the static perfect cosmological fluid case. First, the eqn \eqref{2201d} will simplify as \cite{SSpaper}
\begin{align}
\left(1+\alpha\right)\,\left(\ln\,A_1\right)' +\alpha\,\left(\ln\,\rho\right)' =0 ,  \label{2210}
\end{align}
where $\rho'(r)=\partial_r\,\rho(r)$. By integration, we find as solution for eqn \eqref{2210}:
\begin{align}
\rho(r) = \rho_0\,\left[A_1(r)\right]^{-\frac{\left(1+\alpha\right)}{\alpha}}.  \label{2211}
\end{align}
In a such case, the density of the fluid $\rho(r)$ is directly dependent on $A_1(r)$ for $\alpha \neq 0$ and the energy condition constraints to satisfy $\rho(r)\geq 0$ for positive mass density. For $\alpha=0$ (dust fluid), we will need to solve this case separately for avoiding the singular solution for eqn \eqref{2211}. If we set an ansatz for the $A_i$, $\rho(r)$ will depend directly on this same ansatz according to the conservation laws. But since $\rho(r)$ depends only on $A_1(r)$, one can in principle perform a coordinate change $dt'\,\rightarrow\,A_1(r)\,dt$ for going to a frame where we have a constant and positive fluid density $\rho=\rho_0$ \cite{SSpaper}.

Then, although the eqn \eqref{2201a} remains unchanged, the eqns \eqref{2202b} and \eqref{2202c} will simplify as follows:
\begin{subequations}
\begin{align}
\kappa\,\rho =& \frac{2\,F'(T)}{\left(1+\alpha\right)}\,\left[-\frac{g_1(r)\,k_2(r)}{k_1(r)}+g_2(r)\right] , \label{2212b}
\\
\kappa\,\rho =& -\frac{F(T)}{2} + 2\,F'(T)\,\left[-\frac{g_1(r)\,k_2(r)}{k_1(r)}+g_3(r)\right] . \label{2212c}
\end{align}
\end{subequations} 
With the eqns \eqref{2212b} and \eqref{2212c}, we can put them together eliminating $\rho$ to finally have a relation linking $F(T)$ and $F'(T)$:
\begin{align}
F(T) =& 4\,F'(T)\,\left[-\frac{\alpha}{1+\alpha}\,\frac{g_1(r)\,k_2(r)}{k_1(r)}+\left(g_3(r)-\frac{g_2(r)}{1+\alpha}\right)\right] . \label{2213}
\end{align}
The torsion scalar is:
\begin{align}\label{2213b}
T(r) = -2\left(\frac{\delta}{A_3}+\frac{A_3'}{A_2\,A_3}\right)\left(\frac{\delta}{A_3}+\frac{A_3'}{A_2\,A_3}+\frac{2\,A_1'}{A_1\,A_2}\right).
\end{align}

There are a number of possible approaches for solutions to the FEs described by eqns \eqref{2201a}, \eqref{2212b}, \eqref{2212c} and \eqref{2213b} added by conservation law solution described by eqn \eqref{2211}. The main goal is to find several possible $F(T)$ solutions from these previous equations. For this purpose, we will solve for $A_3=$ constant and $A_3=r$ as in ref \cite{SSpaper}. We can do this because there is a set of coordinates where $A_3=r$ is valuable without any lost of generality and the constant $A_3$ system is the exception to this rule. This consideration is only for a locally coordinate definition. The constant $A_3$ case is an exception because we cannot perform a local transformation allowing to change this into a non-constant term. All other non-constant $A_3$ can be changed by a local transformation into a $A_3=r$ system.


\subsection{Constant $A_3$ field equation solutions}\label{sect31}

By setting $A_3=c_0=$ constant in our FEs, eqns \eqref{2201a}, \eqref{2212b} and \eqref{2212c} become with eqns \eqref{2132i} components:
\small
\begin{subequations}
\begin{align}
F'(T)=& F'(0)\,\exp\Bigg[-\int\,dr\,\frac{\left(A_1''-\frac{A_1'\,A_2'}{A_2}+\frac{A_1\,A_2^2}{c_0^2}\right)}{\left(A_1'+\frac{\delta\,A_1\,A_2}{c_0}\right)}\Bigg] , \label{2214a}
\\
\kappa\,\left(1+\alpha\right)\,\rho =& 2\,F'(T)\,\Bigg[\frac{\left(A_1''-\frac{A_1'\,A_2'}{A_2}+\frac{A_1\,A_2^2}{c_0^2}\right)}{\left(A_1'+\frac{\delta\,A_1\,A_2}{c_0}\right)}\,\frac{\delta}{A_2\,c_0}\Bigg] , \label{2214b}
\\
\kappa\,\rho =& -\frac{F(T)}{2} + 2\,F'(T)\,\Bigg[\frac{\left(A_1''-\frac{A_1'\,A_2'}{A_2}+\frac{A_1\,A_2^2}{c_0^2}\right)}{\left(A_1'+\frac{\delta\,A_1\,A_2}{c_0}\right)}\,\frac{\delta}{A_2\,c_0}-\frac{\delta\,A_1'}{A_1\,A_2\,c_0}\Bigg] . \label{2214c}
\end{align}
\end{subequations} 
\normalsize
Eqn \eqref{2213b} for torsion scalar becomes:
\begin{align}\label{2214d}
T(r) = -\frac{2}{c_0^2}-\frac{4\delta\,A_1'}{c_0\,A_1\,A_2}.
\end{align}

\subsubsection{Power-law solutions}\label{sect311}

We will solve eqns \eqref{2214a} to \eqref{2214c} by using the following ansatz:
\begin{align}\label{2160a}
A_1(r)=a_0\,r^a , \quad\quad\quad A_2(r)=b_0\,r^b . 
\end{align}
In supplement, eqn \eqref{2211} from conservation laws becomes:
\begin{align}
\rho(r) = \rho_1\,r^{-\frac{a\left(1+\alpha\right)}{\alpha}}.  \label{2230e}
\end{align}
where $\rho_1=\rho_0\,a_0^{-\frac{a\left(1+\alpha\right)}{\alpha}}$ and $\alpha \neq 0$. Eqns \eqref{2214a} to \eqref{2214d} become:
\small
\begin{subequations}
\begin{align}
F'(T)=& F'(0)\,\exp\Bigg[\int\,dr\,\frac{\left[a(1-a+b)\,r^{-2(b+1)}-\left(\frac{b_0}{c_0}\right)^2\right]}{r\,\left[a\,r^{-2(b+1)}+\delta\,\left(\frac{b_0}{c_0}\right)\,r^{-(b+1)}\right]}\Bigg] \label{2230b}
\\
\kappa\,\rho(r) =& -\frac{2\delta\,F'(T(r))}{\left(1+\alpha\right)\,b_0c_0}\,\Bigg[\frac{\left[a(1-a+b)\,r^{-2(b+1)}-\left(\frac{b_0}{c_0}\right)^2\right]}{\left[a\,r^{-(b+1)}+\delta\,\left(\frac{b_0}{c_0}\right)\right]}\Bigg] , \label{2230c}
\\
\kappa\,\rho(r) =& -\frac{F(T(r))}{2} -\frac{2\delta\,F'(T(r))}{b_0c_0}\,\Bigg[a\,r^{-(b+1)}+\frac{\left[a(1-a+b)\,r^{-2(b+1)}-\left(\frac{b_0}{c_0}\right)^2\right]}{\left[a\,r^{-(b+1)}+\delta\,\left(\frac{b_0}{c_0}\right)\right]}\Bigg] . \label{2230d}
\\
T(r) =& -\frac{2}{c_0^2}-\frac{4\delta a}{b_0\,c_0}\,r^{-(b+1)} \label{2230a}
\end{align}
\end{subequations} 
\normalsize
For setting eqns \eqref{2230b} to \eqref{2230d} in terms of torsion scalar $T$, we isolate $r(T)$ from eqn \eqref{2230a}:
\begin{align}
r^{-(b+1)}(T)=& -\frac{\delta\,b_0\,c_0}{4a} \left(T+\frac{2}{c_0^2}\right) \label{2231}
\end{align}
By substituting eqn \eqref{2231} into eqns \eqref{2230b} to \eqref{2230d} and by simplifying eqn \eqref{2230b}, we obtain:
\begin{subequations}
\begin{align}
F'(T)=& F'(0)\,\left(T-\frac{2}{c_0^2}\right)^{\frac{2a}{(1+b)}-1}\left(T+\frac{2}{c_0^2}\right)^{-\frac{a}{(b+1)}}\exp\Bigg[\frac{4a}{c_0^2\,(b+1)\left(T+\frac{2}{c_0^2}\right)}\Bigg] \label{2231b}
\\
\kappa\,\rho =& \frac{2\,F'(T)}{\left(1+\alpha\right)}\,\Bigg[\frac{\frac{(1+b-a)}{4a}\,\left(T+\frac{2}{c_0^2}\right)^2-\frac{4}{c_0^4}}{\left(T-\frac{2}{c_0^2}\right)}\Bigg] , \label{2231c}
\\
\kappa\,\rho =& -\frac{F(T)}{2} +2\,F'(T)\,\Bigg[\frac{1}{4} \left(T+\frac{2}{c_0^2}\right)+\frac{\frac{(1+b-a)}{4a}\,\left(T+\frac{2}{c_0^2}\right)^2-\frac{4}{c_0^4}}{\left(T-\frac{2}{c_0^2}\right)}\Bigg] . \label{2231d}
\end{align}
\end{subequations} 
By putting eqns \eqref{2231c} and \eqref{2231d} and then by substituting eqn \eqref{2231b}, we find as solution for $F(T)$:
\begin{align}
F(T)=&\,F'(0)\left(T-\frac{2}{c_0^2}\right)^{\frac{2a}{(1+b)}-2}\left(T+\frac{2}{c_0^2}\right)^{-\frac{a}{(b+1)}}\exp\Bigg[\frac{4a}{c_0^2\,(b+1)\left(T+\frac{2}{c_0^2}\right)}\Bigg]
\nonumber\\
&\quad\times\,\Bigg[\left(T+\frac{2}{c_0^2}\right)\left(T-\frac{2}{c_0^2}\right)+\frac{\alpha}{\left(1+\alpha\right)}\left(\frac{(1+b-a)}{a}\,\left(T+\frac{2}{c_0^2}\right)^2-\frac{16}{c_0^4}\right)\Bigg] , \label{2231e}
\end{align}
where $a\neq 0$ and $b\neq -1$. The eqn \eqref{2231e} is a new non-trivial $F(T)$ teleparallel solution arising from $A_3=$ constant case. Then eqn \eqref{2230e} for the fluid density in terms of $T$ will be expressed as:
\begin{align}
\rho(T) = \rho_2\,\left(T+\frac{2}{c_0^2}\right)^{-\frac{a\left(1+\alpha\right)}{(1+b)\alpha}}.  \label{2231f}
\end{align}
where $\rho_2=\rho_0\,a_0^{-\frac{a\left(1+\alpha\right)}{\alpha}}\left(-\frac{4\delta\,a}{b_0\,c_0}\right)^{\frac{a\left(1+\alpha\right)}{(1+b)\alpha}}$. Therefore eqn \eqref{2231e} has two possible singularities:
\begin{itemize}
\item $T=-\frac{2}{c_0^2}$: This singularity appears in two terms of eqn \eqref{2231e} leading to an undefined $\lim_{T\to -\frac{2}{c_0^2}} F\left(T\right)$ and then $F(T)$ is undefined in all situations. For fluid density, eqn \eqref{2231f} will lead to the following situations:
\begin{itemize}
\item $\frac{a}{(1+b)\alpha}>0$ subcase: $\rho(T)$ is undefined.

\item $\frac{a}{(1+b)\alpha}<0$ subcase: $\rho(T)=0$, the vacuum situation.
\end{itemize}
Then eqn \eqref{2231} will lead to the situations:
\begin{itemize}
\item $b>-1$ subcase: $r(T)$ is undefined.

\item $b<-1$ subcase: $r(T)\,\rightarrow\,0$: a point-like singularity. 
\end{itemize}

\item $T=+\frac{2}{c_0^2}$: This singularity only occurs for $b \neq -1$ and $a<1+b$. For eqns \eqref{2231} and \eqref{2231f}, there is no real consequences because we obtain definite values of $r(T)$ and $\rho(T)$. This is only that $ \lim_{T\to +\frac{2}{c_0^2}} F\left(T\right)\,=\,\infty$.
\end{itemize}

\noindent For $a=b+1$: eqn \eqref{2231e} becomes:
\begin{align}
F(T)=&\,F'(0)\,\exp\Bigg[\frac{4}{c_0^2\,T+2}\Bigg]\,\Bigg[\left(T-\frac{2}{c_0^2}\right)-\frac{16\alpha}{c_0^4\left(1+\alpha\right)}\left(T+\frac{2}{c_0^2}\right)^{-1}\Bigg] , \label{2231g}
\end{align}
and then eqn \eqref{2231f} will simplify as:
\begin{align}
\rho(T) = \rho_2\,\left(T+\frac{2}{c_0^2}\right)^{-\frac{\left(1+\alpha\right)}{\alpha}},  \label{2231h}
\end{align}
where $\rho_2=\rho_0\,a_0^{-\frac{a\left(1+\alpha\right)}{\alpha}}\left(-\frac{4\delta\,a}{b_0\,c_0}\right)^{\frac{\left(1+\alpha\right)}{\alpha}}$. The $T=-\frac{2}{c_0^2}$ singularity is now the remaining one inside eqn \eqref{2231g} and leads to an undefined $F(T)$. We obtain 
from eqn \eqref{2231h} that the fluid density:
\begin{itemize}
\item $\alpha > 0$ subcase: $\rho(T)$ is undefined.

\item $\alpha < 0$ subcase: $\rho(T)=0$, the vacuum situation.
\end{itemize}
For eqn \eqref{2231}, we find that: 
\begin{itemize}
\item $b>-1$ subcase: $r(T)$ is undefined.

\item $b<-1$ subcase: $r(T)\,\rightarrow\,0$, a point-like singularity.
\end{itemize}

\noindent For $a=0$ and/or $b=-1$: these are constant torsion scalar spacetimes cases according to eq \eqref{2230a} and are GR solutions.


\subsubsection{Constant $A_2$ and exponential $A_1$ solutions}\label{sect312}

Another possible ansatz for $F(T)$ solutions is $A_1(r)=a_0\left(1-e^{-ar}\right)$ and $A_2(r)=b_0=1$. We replace the component $A_1$ of the simple power-law ansatz expressed in eqn. \eqref{2160a} by an infinite series of power-laws leading to an exponential ansatz defined as $A_1(r)=a_0\,\sum_{k\geq 0}\,\frac{(-1)^{k}}{(k+1)!}\,(a\,r)^{k+1}$. We then set $b=0$ for the component $A_2$ of this same ansatz thus generalizing the power-law ansatz as expressed in eqn \eqref{2160a}. Then eqn \eqref{2214d} becomes:
\begin{align}
T(r) =& -\frac{2}{c_0^2}-\frac{4\delta\,a}{c_0\,\left(e^{ar}-1\right)}  
\nonumber\\
&\Rightarrow\; e^{-a\,r(T)} = \frac{\left(T+\frac{2}{c_0^2}\right)}{\left(T+\frac{2}{c_0^2}\left(1-2\delta\,a\,c_0\right)\right)} \label{2216d}
\end{align}
The eqns \eqref{2214a} to \eqref{2214c} become:
\small
\begin{subequations}
\begin{align}
F'(T)=&  F'(0)\,\left(T-\frac{2}{c_0^2}\right)^{\frac{\left(\delta+a\,c_0\right)}{(\delta-a\,c_0)}}\left(T+\frac{2}{c_0^2}\right)^{\frac{\delta}{a\,c_0}}\left(T+\frac{2}{c_0^2}\left(1-2\delta\,a\,c_0\right)\right)^{-\frac{\left(1+a^2\,c_0^2\right)}{a\,c_0(\delta-a\,c_0)}} , \label{2216a}
\\
-\kappa\,\rho =& \frac{F'(T)}{\left(1+\alpha\right)}\,\Bigg[\frac{\frac{8}{c_0^4}+\frac{2\delta a}{c_0}\,\left(T+\frac{2}{c_0^2}\right)}{\left(T-\frac{2}{c_0^2}\right)}\Bigg] , \label{2216b}
\\
-\kappa\,\rho =& \frac{F(T)}{2} + F'(T)\,\Bigg[\frac{\left(\frac{8}{c_0^4}+\frac{2\delta a}{c_0}\,\left(T+\frac{2}{c_0^2}\right)\right)}{\left(T-\frac{2}{c_0^2}\right)}-\frac{\left(T+\frac{2}{c_0^2}\right)}{2}\Bigg] . \label{2216c}
\end{align}
\end{subequations} 
\normalsize
By putting eqns \eqref{2216b} and \eqref{2216c} together and then by substituting eqn \eqref{2216a}, we find that:
\begin{align}
F(T) =&  F'(0)\,\left(T-\frac{2}{c_0^2}\right)^{\frac{\left(\delta+a\,c_0\right)}{(\delta-a\,c_0)}}\left(T+\frac{2}{c_0^2}\right)^{\frac{\delta}{a\,c_0}}\left(T+\frac{2}{c_0^2}\left(1-2\delta\,a\,c_0\right)\right)^{-\frac{\left(1+a^2\,c_0^2\right)}{a\,c_0(\delta-a\,c_0)}}\,
\nonumber\\
&\quad\times\;\Bigg[-\frac{2\alpha}{\left(1+\alpha\right)}\frac{\left(\frac{8}{c_0^4}+\frac{2\delta a}{c_0}\,\left(T+\frac{2}{c_0^2}\right)\right)}{\left(T-\frac{2}{c_0^2}\right)}+\left(T+\frac{2}{c_0^2}\right)\Bigg] , \label{2216e}
\end{align}
where $a \neq \left\lbrace 0,\,\frac{\delta}{c_0} \right\rbrace$. The eqn \eqref{2216e} is another new non-trivial $F(T)$ teleparallel solution with $A_3=$ constant. Then eqn \eqref{2211} for the fluid density in terms of $T$ will be:
\begin{align}
\rho(T) = \rho_3\,\left(T+\frac{2}{c_0^2}\left(1-2\delta\,a\,c_0\right)\right)^{\frac{\left(1+\alpha\right)}{\alpha}},  \label{2216f}
\end{align}
where $\rho_3=\rho_0\,\left(-\frac{4\delta\,a_0\,a}{c_0}\right)^{-\frac{\left(1+\alpha\right)}{\alpha}}$ and $\alpha \neq 0$. From eqn \eqref{2216e}, we find $3$ singularities:
\begin{itemize}
\item $T=-\frac{2}{c_0^2}$: This singularity arises when $\frac{\delta}{a\,c_0}<0$ and $\lim_{T\to -\frac{2}{c_0^2}} F\left(T\right)$ is undefined. There is no impact on $\rho(T)$, only a limit of $\rho(T)=\rho_0$. We have that $r(T)\,\rightarrow\,\infty$ for $a>0$ and $r(T)$ is undefined for $a<0$ according to eqn \eqref{2216d}.

\item $T=\frac{2}{c_0^2}$: This singularity arises when $ac_0<0$ only and leads to an undefined $\lim_{T\to \frac{2}{c_0^2}} F\left(T\right)$. There is no real consequences on $\rho(T)$ and $r(T)$: these quantities will be constant.

\item $T=\frac{2}{c_0^2}\left(2\delta\,a\,c_0-1\right)$: this case arises when $0 < a\,c_0 < 1$ for $\delta=+1$ and $-1 < a\,c_0 < 0$ for $\delta=-1$ only, all leading to an undefined $\lim_{T\to \frac{2}{c_0^2}\left(2\delta\,a\,c_0-1\right)} F\left(T\right)$, an undefined $\rho(T)$ for $\alpha<0$ and to $\rho(T)=0$ (vacuum case) for $\alpha>0$ according to eqn \eqref{2216f}. For eqn \eqref{2216d}, we find that $r(T)\,\rightarrow\,\infty$ if $a<0$ (point-like singularity) and $r(T)$ is undefined if $a>0$.
\end{itemize}

\noindent For $a=\frac{\delta}{c_0}$, eqn \eqref{2216d} will be:
\begin{align}
e^{-\frac{\delta}{c_0}\,r(T)} = \frac{\left(T+\frac{2}{c_0^2}\right)}{\left(T-\frac{2}{c_0^2}\right)}, \label{2217d}
\end{align}
where $T \neq \frac{2}{c_0^2}$. Then eqns \eqref{2216a} to \eqref{2216c} will become:
\small
\begin{subequations}
\begin{align}
F'(T)=& F'(0)\,\frac{\left(T+\frac{2}{c_0^2}\right)}{\left(T-\frac{2}{c_0^2}\right)}\exp\Bigg[-2\,\frac{\left(T+\frac{2}{c_0^2}\right)}{\left(T-\frac{2}{c_0^2}\right)}\Bigg] , \label{2217a}
\\
-\kappa\,\rho =& \frac{2F'(T)}{c_0^2\left(1+\alpha\right)}\,\frac{\left(T+\frac{6}{c_0^2}\right)}{\left(T-\frac{2}{c_0^2}\right)} , \label{2217b}
\\
-\kappa\,\rho =& \frac{F(T)}{2} + \frac{2F'(T)}{c_0^2}\,\Bigg[\frac{\left(T+\frac{6}{c_0^2}\right)}{\left(T-\frac{2}{c_0^2}\right)}-\frac{c_0^2}{4}\left(T+\frac{2}{c_0^2}\right)\Bigg] . \label{2217c}
\end{align}
\end{subequations} 
\normalsize
By putting eqns \eqref{2217b} to \eqref{2217c} and then by substituting eqn \eqref{2217a} inside, we find that:
\small
\begin{align}
F(T)=&\frac{F'(0)}{c_0^2}\,\frac{\left(T+\frac{2}{c_0^2}\right)}{\left(T-\frac{2}{c_0^2}\right)^2}\exp\Bigg[-2\,\frac{\left(T+\frac{2}{c_0^2}\right)}{\left(T-\frac{2}{c_0^2}\right)}\Bigg]\Bigg[c_0^2\,\left(T+\frac{2}{c_0^2}\right)\left(T-\frac{2}{c_0^2}\right)-\frac{4\alpha}{\left(1+\alpha\right)}\left(T+\frac{6}{c_0^2}\right)\Bigg]  . \label{2217e}
\end{align}
\normalsize
Then eqn \eqref{2216f} for fluid density will be:
\begin{align}
\rho(T) = \rho_3\,\left(T-\frac{2}{c_0^2}\right)^{\frac{\left(1+\alpha\right)}{\alpha}},  \label{2217f}
\end{align}
where $\rho_3=\rho_0\,\left(-\frac{4\,a_0}{c_0^2}\right)^{-\frac{\left(1+\alpha\right)}{\alpha}}$ and $\alpha \neq 0$. In the case of eqn \eqref{2217e}, the only and remaining singularity is $T=\frac{2}{c_0^2}$ leading to an undefined $F(T)$, $\rho(T)=0$ for $\alpha>0$ (vacuum) and an undefined $\rho(T)$ for $\alpha<0$ according to eqn \eqref{2217f}, all with $r\,\rightarrow\,\infty$ and $\frac{\delta}{c_0}<0$ from eqn \eqref{2217d}.

\noindent For $a=0$, we obtain from eqn \eqref{2216d} that the torsion scalar is constant (i.e. $T=-\frac{2}{c_0^2}$), $A_1=a_0$ and $\rho(T)=\rho_0\,a_0^{-\frac{\left(1+\alpha\right)}{\alpha}}=$ constant leading to GR solutions.

In comparison with ref \cite{SSpaper}, we obtain as result for the pure vacuum case a linear $F(T)$, which is a GR solution. But for a perfect fluid with $\alpha \neq 0$, we find some new and non-trivial teleparallel $F(T)$ specific solutions. These are all new teleparallel fluid non-vacuum solutions for $A_3=$ constant class.


\subsection{$A_3=r$ field equation solutions}\label{sect32}

For $A_3=r$ FEs, eqns \eqref{2201a}, \eqref{2212b} and \eqref{2212c} become with eqns \eqref{2923i} components:
\small
\begin{subequations}
\begin{align}
F'(T)=& F'(0)\,\exp\Bigg[\int\,\frac{dr}{A_2\,r}\,\frac{\left[-A_2\,r^2\,A_1''+A_1\,A_2+\left(A_1\,A_2\right)'\,r+r^2\,A_1'\,A_2'-A_1\,A_2^3\right]}{\left[A_1+r\,A_1'+\delta\,A_1\,A_2\right] }\Bigg] , \label{2224a}
\\
\kappa\,\left(1+\alpha\right)\,\rho =& 2\,F'(T)\,\Bigg[-\frac{\left[-A_2\,r^2\,A_1''+A_1\,A_2+\left(A_1\,A_2\right)'\,r+r^2\,A_1'\,A_2'-A_1\,A_2^3\right]}{\left[A_1+r\,A_1'+\delta\,A_1\,A_2\right]} \,\frac{\left(1+\delta\,A_2\right)}{A_2^3\,r^2}
\nonumber\\
&\quad+\frac{\left(A_1\,A_2\right)'}{A_1\,A_2^3\,r}\Bigg] , \label{2224b}
\\
\kappa\,\rho+\frac{F(T)}{2} =& 2\,F'(T)\,\Bigg[-\frac{\left[-A_2\,r^2\,A_1''+A_1\,A_2+\left(A_1\,A_2\right)'\,r+r^2\,A_1'\,A_2'-A_1\,A_2^3\right]}{\left[A_1+r\,A_1'+\delta\,A_1\,A_2\right] }\,\frac{\left(1+\delta\,A_2\right)}{A_2^3\,r^2}
\nonumber\\
&+\frac{1}{A_1\,A_2^3\,r^2}\left[-A_1\,A_2-A_2\,r\,A_1'-\delta\,A_1\,A_2^2+A_1\,r\,A_2'-\delta\,A_2^2\,r\,A_1'\right]\Bigg] . \label{2224c}
\end{align}
\end{subequations} 
\normalsize
Eqn \eqref{2213b} for torsion scalar becomes:
\begin{eqnarray}\label{2224d}
T(r)=& -\frac{2}{r^2\,A_2^2}\left[\left(\delta\,A_2+1\right)^2+\frac{2\,r\,A_1'}{A_1}\,\left(\delta\,A_2+1\right)\right].
\end{eqnarray} 
There are a number of approach for solving eqns \eqref{2224a} to \eqref{2224d} to find specific pure $F(T)$ new solutions in general perfect fluid case with $\alpha \neq 0$. For conservation laws, $\rho(r)$ is still described by eqn. \eqref{2211}, because $\rho(r)$ depends only on the $A_1(r)$ component.

\subsubsection{General power-law field equations}\label{sect321}

For FEs and conservation law in terms of power-law solutions, we will use the eqn \eqref{2160a} ansatz in eqns \eqref{2230e} and \eqref{2224a} to \eqref{2224d}. From there, we obtain:
\small
\begin{subequations}
\begin{align}
F'(T(r))=& F'(T(0))\,\exp\left[\int\,dr\,\frac{\left[\left(2a-a^2+ab+b+1\right)\,r^{-2b}-b_0^2\right]}{\left[(a+1)\,r^{1-2b}+\delta\,b_0\,r^{1-b}\right]}\right] \label{2225a}
\\
\kappa\,\left(1+\alpha\right)\,\rho =& \frac{2\,F'(T)}{b_0^2}\,\left[-\frac{\left[\left(2a-a^2+ab+b+1\right)\,r^{-2b}-b_0^2\right]}{\left[(a+1)\,r^{1-2b}+\delta\,b_0\,r^{1-b}\right]}\,\Bigg[r^{-2b-1}+\delta\,b_0\,r^{-b-1}\Bigg]+(a+b)\,r^{-2b-2}\right] , \label{2225b}
\\
\kappa\,\rho =& -\frac{F(T)}{2} + \frac{2\,F'(T)}{b_0^2}\,\Bigg[-\frac{\left[\left(2a-a^2+ab+b+1\right)\,r^{-2b}-b_0^2\right]}{\left[(a+1)\,r^{2(1-b)-1}+\delta\,b_0\,r^{(1-b)}\right]}\,\Bigg[r^{-2b-1}+\delta\,b_0\,r^{-b-1}\Bigg]
\nonumber\\
&\quad+(-a+b-1)\,r^{-2b-2}-\delta\,b_0\,(a+1)\,r^{-b-2}\Bigg], \label{2225c}
\\
T(r)=& -\frac{2}{b_0^2}\left[b_0^2\,r^{-2}+2\delta\,b_0\,(1+a)\,r^{-2-b}+(2\,a+1)\,r^{-2-2b}\right], \label{2225d}
\\
\rho(r)=&\rho_1\,r^{-\frac{a\,\left(1+\alpha\right)}{\alpha}} ,\label{2225f}
\end{align}
\end{subequations} 
\normalsize
where $\rho_1=a_0^{-\frac{a\left(1+\alpha\right)}{\alpha}}=$ constant and $\alpha \neq 0$. From eqn \eqref{2225d}, we find the following characteristic equation for $r(T)$:
\begin{align}\label{2225e}
0=& \frac{b_0^2\,T}{2}+b_0^2\,r^{-2}+2\delta\,b_0\,(1+a)\,r^{-2-b}+(2\,a+1)\,r^{-2-2b}.
\end{align}
From this eqn \eqref{2225e}, we can in principle isolate for each value of $a$ and $b$ a relation $r(T)$ for finding specific solution $F(T)$, which is the main aim of this rigourous works.

\subsubsection{Simple spacetime solutions}\label{sect322}

Before going to more complex solutions, it is important to consider the simplest case of pure flat cosmological spacetimes where ${\bf a=b=0}$. In this case, eqn \eqref{2225e} becomes:
\begin{align}
0=& \frac{b_0^2\,T}{2}+\left(1+\delta\,b_0\right)^2\,r^{-2},
\nonumber\\
&\Rightarrow\,r^{-2}(T)=\frac{b_0^2}{2\left(1+\delta\,b_0\right)^2} (-T). \label{2380d}
\end{align}
Then eqns \eqref{2225a} to \eqref{2225c} and eqn \eqref{2230e} become:
\small
\begin{subequations}
\begin{align}
F'(T)=& F'(0)\,\left(\frac{b_0}{\sqrt{2}\left(1+\delta\,b_0\right)}\right)^{\delta b_0-1}\,(-T)^{\frac{\delta b_0-1}{2}} \label{2380a}
\\
\kappa\,\rho =& \frac{\left(1-\delta b_0\right)}{\left(1+\delta\,b_0\right)\left(1+\alpha\right)}\,T\,F'(T) , \label{2380b}
\\
\kappa\,\rho =& -\frac{F(T)}{2} +\frac{\left(2-\delta b_0\right)}{\left(1+\delta\,b_0\right)}\,T\,F'(T), \label{2380c}
\\
\rho=&\rho_1=\rho_0\,a_0^{-\frac{a\left(1+\alpha\right)}{\alpha}}=\text{const.}  \label{2380f}
\end{align}
\end{subequations} 
\normalsize
Eqns \eqref{2380a} to \eqref{2380c} is expressed in terms of $T$, $F(T)$ and $F'(T)$ only. By putting eqns \eqref{2380b} and \eqref{2380c} together, we find that:
\begin{align}
F(T)=F(0)\,T^{\frac{\left(1+\delta\,b_0\right)\left(1+\alpha\right)}{2\left(1+\alpha\left(2-\delta b_0\right)\right)}} , \label{2380e}
\end{align}
where $F(0)$ is an integration constant. Eqn \eqref{2380e} is a pure power-law solution for static simple cosmological spacetimes where $\alpha \neq 0$, which is similar to Bahamonde-Camci solutions \cite{baha1}. If $b_0=\delta$, we find that eqn \eqref{2380e} will be reduced to the TEGR-like solution $F(T)=F(0)\,T$. For a pure flat null torsion spacetime, we need that $b_0=-\delta$ in eqns \eqref{2380d} to \eqref{2380c}, which is leading to $T=0$ and $F(T)=F(0)=$ constant without any other condition. If then $F(0)=0$, we obtain the pure Minkowski spacetime \cite{landryvandenhoogen1}. 


\subsubsection{General case solutions}\label{sect323}

\noindent We will consider different cases according to value of $b$ for the general case (i.e. $a \neq \left\lbrace -1,\,-\frac{1}{2}\right\rbrace $):
\begin{enumerate}
\item ${\bf b=0}$ case: Eqn \eqref{2225e} will be:
\begin{align}\label{2226d}
0=& \frac{b_0^2\,T}{2}+\left(b_0^2+2\delta\,b_0\,(1+a)+2\,a+1\right)\,r^{-2},
\nonumber\\
&\quad\Rightarrow\,r^{-1}(T)=\pm\,\frac{b_0}{\sqrt{2}\,\sqrt{b_0^2+2\delta\,b_0\,(1+a)+2\,a+1}}\,\left(-T\right)^{1/2},
\end{align}
where $T \leq 0$. As for the simple case presented in section \ref{sect322}, we substitute eqn \eqref{2226d} into eqns \eqref{2225a} to \eqref{2225c} by setting $b=0$. After that, by putting eqns \eqref{2225b} and \eqref{2225c} together, and then by substituting eqn \eqref{2225a}, we find:
\begin{align}
F(T) =&\,4F'(0)\left(2\,\left(b_0^2+2\delta\,b_0\,(1+a)+2\,a+1\right)\right)^{-\frac{\left(a^2+\left(1+\delta\,b_0\right)^2\right]}{2\left(a+1+\delta\,b_0\right)}}\,b_0^{-\frac{\left[2a-a^2+1-b_0^2\right]}{\left(a+1+\delta\,b_0\right)}}
\nonumber\\
&\;\times\,\Bigg[\frac{\alpha\,\left[a^2-2a-1+b_0^2\right]}{\left(1+\alpha\right)\left(a+1+\delta\,b_0\right)}\,\left(1+\delta\,b_0\right)-\frac{a}{\left(1+\alpha\right)}-(a+1)\left( 1+\delta\,b_0\right)\Bigg]
\nonumber\\
&\;\times\,\left(-T\right)^{\frac{\left(a^2+\left(1+\delta\,b_0\right)^2\right]}{2\left(a+1+\delta\,b_0\right)}},
\nonumber\\
=& F_1\,\left(-T\right)^{\frac{\left(a^2+\left(1+\delta\,b_0\right)^2\right]}{2\left(a+1+\delta\,b_0\right)}}, \label{2226e}
\end{align}
where $F_1$ is a constant. Here we have a power-law solution similar to the Bahamonde-Camci solution where $\rho(r)$ is described by eqn \eqref{2230e} \cite{baha1}. In terms of torsion scalar, this eqn \eqref{2230e} becomes:
\begin{align}
\rho(T) = \rho_1\,\left(\frac{b_0^2}{2\,\left(b_0^2+2\delta\,b_0\,(1+a)+2\,a+1\right)}\right)^{\frac{a\left(1+\alpha\right)}{2\alpha}}\,\left(-T\right)^{\frac{a\left(1+\alpha\right)}{2\alpha}},  \label{2226f}
\end{align}
where $\alpha \neq 0$. We have a direct density-linked $F(T)$ solution in this case.


\item ${\bf b=1}$ case: Eqn \eqref{2225e} will be:
\begin{align}\label{2235}
0=& \frac{b_0^2\,T}{2}+b_0^2\,r^{-2}+2\delta\,b_0\,(1+a)\,r^{-3}+(2\,a+1)\,r^{-4},
\end{align}
The solutions are:
\small
\begin{align}
&r^{-1}(T)=
\nonumber\\
&\-\frac{\delta_1\,b_0(a+1)}{2(2a+1)}-\frac{\delta_1}{2}\Bigg[\Bigg(-\frac{2b_0^2}{3(2a+1)}+\frac{b_0^2(a+1)^2}{(2a+1)^2}
\nonumber\\
&\;+\frac{1}{6\sqrt[3]{2}(2a+1)}\Bigg(\Bigg(16b_0^6+\left(432(a+1)^2-288\,(2a+1)\right)\,b_0^4\,T
\nonumber\\
&\;+\sqrt{\left(\left(16b_0^6+\left(432(a+1)^2-288(2a+1)\right)\,b_0^4\,T\right)^2-4b_0^6\left(4b_0^2+24(2a+1)\,T\right)^3\right)}\Bigg)^{1/3} \Bigg)
\nonumber\\
&\;+\frac{2^{4/3}b_0^2\left(b_0^2+6\,(2a+1)\,T\right)}{3(2a+1)}\Bigg(16b_0^6+\left(432(a+1)^2-288\,(2a+1)\right)\,b_0^4\,T
\nonumber\\
&\;+\sqrt{\left(\left(16b_0^6+\left(432(a+1)^2-288(2a+1)\right)\,b_0^4\,T\right)^2-4b_0^6\left(4b_0^2+24(2a+1)\,T\right)^3\right)}\Bigg)^{-1/3}\Bigg)\Bigg]^{1/2}
\nonumber\\
&\;-\frac{\delta_2}{2}\Bigg[\Bigg(-\frac{4b_0^2}{3(2a+1)}+\frac{2b_0^2(a+1)^2}{(2a+1)^2}
\nonumber\\
&\;-\frac{1}{6\sqrt[3]{2}(2a+1)}\Bigg(\Bigg(16b_0^6+\left(432(a+1)^2-288\,(2a+1)\right)\,b_0^4\,T
\nonumber\\
&\;+\sqrt{\left(\left(16b_0^6+\left(432(a+1)^2-288(2a+1)\right)\,b_0^4\,T\right)^2-4b_0^6\left(4b_0^2+24(2a+1)\,T\right)^3\right)}\Bigg)^{1/3} \Bigg)
\nonumber\\
&\;+\frac{2^{4/3}b_0^2\left(b_0^2+6\,(2a+1)\,T\right)}{3(2a+1)}\Bigg(16b_0^6+\left(432(a+1)^2-288\,(2a+1)\right)\,b_0^4\,T
\nonumber\\
&\;+\sqrt{\left(\left(16b_0^6+\left(432(a+1)^2-288(2a+1)\right)\,b_0^4\,T\right)^2-4b_0^6\left(4b_0^2+24(2a+1)\,T\right)^3\right)}\Bigg)^{-1/3}\Bigg)
\nonumber\\
&+\delta_1\,\frac{8\delta\,b_0^3\,a^2\,(a+1)}{(2a+1)^3}\Bigg[4\Bigg[\Bigg(-\frac{2b_0^2}{3(2a+1)}+\frac{b_0^2(a+1)^2}{(2a+1)^2}
\nonumber\\
&+\frac{1}{6\sqrt[3]{2}(2a+1)}\Bigg(\Bigg(16b_0^6+\left(432(a+1)^2-288\,(2a+1)\right)\,b_0^4\,T
\nonumber\\
&\;+\sqrt{\left(\left(16b_0^6+\left(432(a+1)^2-288(2a+1)\right)\,b_0^4\,T\right)^2-4b_0^6\left(4b_0^2+24(2a+1)\,T\right)^3\right)}\Bigg)^{1/3} \Bigg)
\nonumber\\
&\;+\frac{2^{4/3}b_0^2\left(b_0^2+6\,(2a+1)\,T\right)}{3(2a+1)}\Bigg(16b_0^6+\left(432(a+1)^2-288\,(2a+1)\right)\,b_0^4\,T
\nonumber\\
&\;+\sqrt{\left(\left(16b_0^6+\left(432(a+1)^2-288(2a+1)\right)\,b_0^4\,T\right)^2-4b_0^6\left(4b_0^2+24(2a+1)\,T\right)^3\right)}\Bigg)^{-1/3}\Bigg)\Bigg]^{1/2}\Bigg]^{-1}\Bigg]^{1/2}, \label{2235aa}
\end{align}
\normalsize
where $\left(\delta_1,\,\delta_2\right)=\left(\pm 1,\,\pm 1\right)$. By putting eqns \eqref{2225b} and \eqref{2225c} together and then by substituting eqns \eqref{2225a} and \eqref{2235aa}, we find as solution:
\begin{align}
F(T)=&\frac{4\,F'(0)}{b_0^2}\,e^{-\delta\,b_0\,r(T)}\,\left(a+1+\delta\,b_0\,r(T)\right)^{\frac{2a^2}{a+1}-1}\,\left[r(T)\right]^{\frac{2+3a-a^2}{a+1}}
\nonumber\\
&\;\times\,\Bigg[-\frac{\alpha\left[\left(3a-a^2+2\right)\,r^{-2}(T)-b_0^2\right]}{\left(1+\alpha\right)\left[(a+1)\,r^{-1}(T)+\delta\,b_0\right]}\,\left(r^{-1}(T)+\delta\,b_0\right)\,r^{-2}(T)
\nonumber\\
&\;-\frac{(a+1)}{\left(1+\alpha\right)}\,r^{-4}(T)-a\,r^{-4}(T)-\delta\,b_0\,(a+1)\,r^{-3}(T)\Bigg], \label{2235e}
\end{align}
where $r(T)$ is  eqns \eqref{2235aa}.

\item ${\bf b=-1}$ case: Eqn \eqref{2225e} will be:
\begin{align}\label{2236}
0=& \left(\frac{b_0^2\,T}{2}+2\,a+1\right)+2\delta\,b_0\,(1+a)\,r^{-1}+b_0^2\,r^{-2},
\nonumber\\
&\;\Rightarrow\,r^{-1}(T)=-\frac{\delta\,(a+1)}{b_0}\pm \sqrt{\frac{a^2}{b_0^2}-\frac{T}{2}}
\end{align}
By putting eqns \eqref{2225b} and \eqref{2225c} together and then by substituting eqns \eqref{2225a} and \eqref{2236}, we find:
\begin{align}
F(T)=&\frac{4\,F'(0)}{b_0^2}\,\left[r(T)\right]^{a+1}\,\left[(a+1)\,r(T)+\delta\,b_0\right]^{-\frac{2a^2+a+1}{a+1}}\,\exp\left(\frac{\delta\,b_0}{r(T)}\right)
\nonumber\\
&\;\times\,\Bigg[-\frac{\alpha\,\left[a\left(1-a\right)\,r^{2}(T)-b_0^2\right]\left(r(T)+\delta\,b_0\right)}{\left(1+\alpha\right)\left[(a+1)\,r^{3}(T)+\delta\,b_0\,r^{2}(T)\right]}+\frac{(1-a)}{\left(1+\alpha\right)}-(a+2)
\nonumber\\
&\;-\delta\,b_0\,(a+1)\,r^{-1}(T)\Bigg], \label{2236e}
\end{align}
where $r(T)$ is the eqn \eqref{2236}.

\item ${\bf b=-2}$ case: Eqn \eqref{2225e} will be:
\begin{align}\label{2237}
0=& b_0^2+\left(\frac{b_0^2\,T}{2}+2\delta\,b_0\,(1+a)\right)\,r^2+(2\,a+1)\,r^{4},
\nonumber\\
&\;\Rightarrow\,r(T)=\pm\frac{1}{2\sqrt{2a+1}}\,\sqrt{\delta_2\,b_0\sqrt{16a^2+8\delta\,(a+1)\,b_0\,T+b_0^2\,T^2} -4\delta\,b_0(a+1)-b_0^2\,T}.
\end{align}
Then we will set the positive $r(T)$ case and $\delta_2=\pm 1$. By putting eqns \eqref{2225b} and \eqref{2225c} together and then by substituting eqns \eqref{2225a} and \eqref{2237}, we find:
\begin{align}
F(T) =& \frac{4\,F'(0)}{b_0^2}\,\Bigg[\frac{\alpha\,\left[\left(a^2+1\right)\,r^{4}(T)+b_0^2\right]\,\left[r^2(T)+\delta\,b_0\right]}{\left(1+\alpha\right)\,r^2(T)\,\left[(a+1)\,r^2(T)+\delta\,b_0\right]}-\frac{(a+2\alpha)}{\left(1+\alpha\right)}\,r^{2}(T)
\nonumber\\
&\;-(a+1)\left[r^{2}(T)+\delta\,b_0\right]\Bigg]\left[(a+1)r^2(T)+b_0+\delta\right]^{-\frac{(a^2+a+1)}{(a+1)}}r^{(a+1)}(T)\,\exp\left(\frac{\delta\,b_0}{2\,r^2(T)}\right), \label{2237e}
\end{align}
where $r(T)$ is the eqn \eqref{2237}.

\end{enumerate}


\subsubsection{$a=-1$ case solutions}\label{sect324}

\noindent For solving ${\bf a=-1}$ specific cases, the eqn \eqref{2225e} will simplify as:
\begin{align}\label{2240d}
0=& \frac{b_0^2\,T}{2}+b_0^2\,r^{-2}-r^{-2-2b}.
\end{align}
Then, eqns \eqref{2225a} to \eqref{2225f} and \eqref{2225f} will be simplified as:
\small
\begin{subequations}
\begin{align}
F'(T)=& F'(0)\,\exp\left[-\frac{\delta}{b_0}\int\,dr\,\left(2\,r^{-b-1}+b_0^2r^{b-1}\right)\right] \label{2240a}
\\
\kappa\,\rho =& \frac{2\,F'(T)}{b_0^2\,\left(1+\alpha\right)}\,\Bigg[\frac{\delta}{b_0}\left[2\,r^{-b-1}+b_0^2\,r^{b-1}\right]\,\left[r^{-2b-1}+\delta\,b_0\,r^{-b-1}\right]+(b-1)\,r^{-2b-2}\Bigg] , \label{2240b}
\\
\kappa\,\rho =& -\frac{F(T)}{2} + \frac{2\,F'(T)}{b_0^2}\,\Bigg[\frac{\delta}{b_0}\left[2\,r^{-b-1}+b_0^2\,r^{b-1}\right]\,\left[r^{-2b-1}+\delta\,b_0\,r^{-b-1}\right]+b\,r^{-2b-2}\Bigg], \label{2240c}
\\
\rho=&\rho_1\,r^{\frac{\left(1+\alpha\right)}{\alpha}} ,  \label{2240f}
\end{align}
\end{subequations} 
\normalsize
where $\rho_1=\rho_0\,a_0^{\frac{\left(1+\alpha\right)}{\alpha}}$ is a constant.

\noindent From this eqn \eqref{2240d}, there are several new subcases arising and leading new $F(T)$ solutions for eqns \eqref{2240a} to \eqref{2240c}. There subcases are:
\begin{enumerate}
\item ${\bf b=0}$: Eqn \eqref{2240d} becomes:
\begin{align}
0=& \frac{b_0^2\,T}{2}+\left(b_0^2-1\right)\,r^{-2}.
\nonumber\\
r^{-2}(T) =& \left(\frac{b_0^2}{2\left(1-b_0^2\right)}\right)\,T  \label{2244a}
\end{align}
By putting together eqns \eqref{2240b} and \eqref{2240c} and then by substituting eqns \eqref{2240a} and \eqref{2244a}, we obtain a power-law $F(T)$ solution:
\begin{align}\label{2244e}
F(T)=&\Bigg[\frac{4F'(0)}{b_0^2\left(1+\alpha\right)}\,\left(\frac{b_0^2}{2\left(1-b_0^2\right)}\right)^{1+\frac{\delta}{b_0}\left(1+\frac{b_0^2}{2}\right)}\left[\delta\alpha\left(2+b_0^2\right)\,\left(1+\delta\,b_0\right)+b_0\right]\Bigg]\,T^{1+\frac{\delta}{b_0}\left(1+\frac{b_0^2}{2}\right)}
\nonumber\\
=& F_2\,T^{1+\frac{\delta}{b_0}\left(1+\frac{b_0^2}{2}\right)},
\end{align}
where $F_2$ is a constant. The eqn \eqref{2244e} is a pure power-law solution and this is similar to the Bahamonde-Camci solution \cite{baha1}.

\item ${\bf b=\frac{1}{2}}$: Eqn \eqref{2240d} becomes:
\begin{align}\label{2246d}
0=& \frac{b_0^2\,T}{2}+b_0^2\,r^{-2}-r^{-3},
\nonumber\\
\Rightarrow\; r^{-1}(T)=&\frac{1}{3} \Bigg[b_0^2+\frac{2^{2/3}\,b_0^4}{\left[3^{3/2}\sqrt{27b_0^4\,T^2+8b_0^8\,T}+27b_0^2\,T+4b_0^6\right]^{1/3}}
\nonumber\\
&\;+\frac{\left[3^{3/2}\sqrt{27b_0^4\,T^2+8b_0^8\,T}+27b_0^2\,T+4b_0^6\right]^{1/3}}{2^{2/3}}\Bigg].
\end{align}
By putting eqns \eqref{2240b} and \eqref{2240c} together, and then by substituting eqns \eqref{2240a} and \eqref{2246d}, we find that:
\begin{align}
F(T) =& \frac{4\,F'(0)}{b_0^2}\,\Bigg[\frac{\delta\,\alpha}{b_0\left(1+\alpha\right)}\left(2\,r^{-1}(T)+b_0^2\right)\,\left(r^{-\frac{1}{2}}(T)+\delta\,b_0\right)\,r^{-2}(T)+\frac{\left(2+\alpha\right)}{2\left(1+\alpha\right)}\,r^{-3}(T)\Bigg]
\nonumber\\
&\;\times\,\exp\left[\frac{2\delta}{b_0}\left(2\,r^{-\frac{1}{2}}(T)-b_0^2r^{\frac{1}{2}}(T)\right)\right] , \label{2246e}
\end{align}
where $r^{-1}(T)$ is eqn \eqref{2246d}.

\item ${\bf b=-\frac{1}{2}}$: Eqn \eqref{2240d} becomes:
\begin{align}
0=& r^{-2}-\frac{1}{b_0^2}\,r^{-1}+\frac{T}{2}
\nonumber\\
&\Rightarrow\; r^{-1}(T)= \frac{1}{2b_0^2}\left[1 +\delta_1 \sqrt{1-2\,b_0^4\,T}\right] \label{2241a}
\end{align}
where $\delta_1=\pm 1$. By putting eqns \eqref{2240b} and \eqref{2240c} together, and then by substituting eqns \eqref{2240a} and \eqref{2241a}, we obtain that:
\small
\begin{align}
F(T)=&\frac{F'(0)}{(1+\alpha)\,b_0^4}\,\exp\Bigg[\frac{\sqrt{2}\left[-3 +\delta_1 \sqrt{1-2\,b_0^4\,T}\right]}{\left[1 +\delta_1 \sqrt{1-2\,b_0^4\,T}\right]^{1/2}}\Bigg]\,\left[1 +\delta_1 \sqrt{1-2\,b_0^4\,T}\right]^{1/2}
\nonumber\\
& \times\Bigg[\Bigg(2+\alpha \left[4 +\delta_1 \sqrt{1-2\,b_0^4\,T}\right]\Bigg)\left[1 +\delta_1 \sqrt{1-2\,b_0^4\,T}\right]^{1/2}+\sqrt{2}\alpha \left[5 +\delta_1 \sqrt{1-2\,b_0^4\,T}\right]\Bigg] . \label{2242d}
\end{align}
\normalsize

\item ${\bf b=1}$, the eqn \eqref{2240d} becomes:
\begin{align}
0 =& r^{-4}-b_0^2\,r^{-2}-\frac{b_0^2\,T}{2}
\nonumber\\
&\Rightarrow\; r^{-2}(T)= \frac{b_0^2}{2}\left[1 +\delta_1 \sqrt{1+\frac{2\,T}{b_0^2}} \right].\label{2241b}
\end{align}
where $\delta_1=\pm 1$. By putting eqns \eqref{2240b} and \eqref{2240c} together, and then by substituting eqns \eqref{2240a} and \eqref{2241b}, we obtain that:
\small
\begin{align}
F(T)=& F'(0)\,b_0^2\exp\Bigg[\frac{+\delta_1\sqrt{2}\sqrt{1+\frac{2\,T}{b_0^2}}}{\left[1 +\delta_1 \sqrt{1+\frac{2\,T}{b_0^2}} \right]^{1/2}}\Bigg]\left[1 +\delta_1 \sqrt{1+\frac{2\,T}{b_0^2}} \right]
\nonumber\\
&\times\Bigg[1 +\delta_1 \sqrt{1+\frac{2\,T}{b_0^2}} +\frac{\alpha\left[2 +\delta_1 \sqrt{1+\frac{2\,T}{b_0^2}} \right]}{(1+\alpha)}\Bigg[\sqrt{2}\left[1 +\delta_1 \sqrt{1+\frac{2\,T}{b_0^2}} \right]^{1/2}+2\Bigg]\Bigg] . \label{2243d}
\end{align}
\normalsize

\item ${\bf b=-1}$: Eqn \eqref{2240d} becomes:
\begin{align}\label{2248d}
0=& \frac{b_0^2\,T}{2}+b_0^2\,r^{-2}-1,
\nonumber\\
&\Rightarrow\; r^{-1}(T)=\delta_1\,\sqrt{\frac{1}{b_0^2}-\frac{T}{2}},
\end{align}
where $T \leq \frac{2}{b_0^2}$. By putting eqns \eqref{2240b} and \eqref{2240c} together, and then by substituting eqns \eqref{2240a} and \eqref{2248d}, we obtain that:
\begin{align}
F(T) =& \frac{4\,F'(0)}{b_0^2}\,\Bigg[\frac{\delta\,\alpha}{b_0\left(1+\alpha\right)}\left(2+b_0^2\,r^{-2}(T)\right)\left(r(T)+\delta\,b_0\right)+\frac{\left(1-\alpha\right)}{\left(1+\alpha\right)}\Bigg]
\nonumber\\
&\;\times\,\exp\left[-\frac{\delta}{b_0}\left(2\,r(T)-b_0^2r^{-1}(T)\right)\right], \label{2248e}
\end{align}
where $r(T)$ is eqn \eqref{2248d}.

\item ${\bf b=-\frac{3}{2}}$: Eqn \eqref{2240d} becomes:
\begin{align}\label{2249d}
0=& r^{-3}+\frac{T}{2}\,r^{-1}-\frac{1}{b_0^2},
\nonumber\\
\Rightarrow &\; r^{-1}(T)=\frac{1}{6^{2/3}b_0^2}\,\left[\sqrt{6}b_0^4\sqrt{b_0^4\,T^3+54}+18b_0^4\right]^{1/3}
\nonumber\\
&\quad\quad\quad\quad\quad-\frac{b_0^2\,T}{6^{1/3}}\left[\sqrt{6}b_0^4\sqrt{b_0^4\,T^3+54}+18b_0^4\right]^{-1/3}.
\end{align}
By putting eqns \eqref{2240b} and \eqref{2240c} together, and then by substituting eqns \eqref{2240a} and \eqref{2249d}, we obtain that:
\begin{align}
F(T)=& \frac{4\,F'(0)}{b_0^2}\,\Bigg[\frac{\delta\,\alpha}{b_0\left(1+\alpha\right)}\left(2+b_0^2\,r^{-3}(T)\right)\left(r^{\frac{3}{2}}(T)+\delta\,b_0\right)\,r(T)+\frac{r(T)\left(2-3\alpha\right)}{2\left(1+\alpha\right)}\Bigg]
\nonumber\\
&\;\times\,\exp\left[-\frac{2\delta}{3b_0}\left(2\,r^{\frac{3}{2}}(T)-b_0^2r^{-\frac{3}{2}}(T)\right)\right], \label{2249e}
\end{align}
where $r(T)$ is eqn \eqref{2249d}.

\item ${\bf b=2}$: Eqn \eqref{2240d} becomes:
\begin{align}\label{2260d}
0=& \frac{b_0^2\,T}{2}+b_0^2\,r^{-2}-r^{-6},
\nonumber\\
\Rightarrow &\; r^{-1}(T)=\delta_1\,\Bigg[\frac{2^{2/3}\,b_0^2}{3^{1/3}}\left[\sqrt{3}b_0^2\sqrt{27\,T^2-16b_0^2}+9b_0^2\,T\right]^{-\frac{1}{3}}
\nonumber\\
&\quad\quad\quad\quad\quad+\frac{1}{6^{2/3}}\left[\sqrt{3}b_0^2\sqrt{27\,T^2-16b_0^2}+9b_0^2\,T\right]^{\frac{1}{3}}\Bigg]^{\frac{1}{2}},
\end{align}
where $\delta_1=\pm 1$. By putting eqns \eqref{2240b} and \eqref{2240c} together, and then by substituting eqns \eqref{2240a} and \eqref{2260d}, we obtain that:
\begin{align}
F(T)=& \frac{4\,F'(0)}{b_0^2}\,\Bigg[\frac{\alpha\,\delta}{b_0\left(1+\alpha\right)}\left(2+b_0^2\,r^4(T)\right)\left(r^{-2}(T)+\delta\,b_0\right)\,r^{-6}(T)+\frac{\left(1+2\alpha\right)}{\left(1+\alpha\right)} r^{-6}(T)\Bigg]
\nonumber\\
&\;\times\,\exp\left[-\frac{\delta}{2b_0}\,\left(-2\,r^{-2}(T)+b_0^2r^2(T)\right)\right], \label{2260e}
\end{align}
where $r(T)$ is eqn \eqref{2260d}.

\item ${\bf b=-2}$: Eqn \eqref{2240d} becomes:
\begin{align}\label{2261d}
0=& \frac{b_0^2\,T}{2}\,r^2+b_0^2-r^{4},
\nonumber\\
&\Rightarrow\; r(T)=\frac{\delta_2}{2} \sqrt{b_0^2\,T+\delta_1\,b_0\sqrt{b_0^2\,T^2+16}}.
\end{align}
By putting eqns \eqref{2240b} and \eqref{2240c} together, and then by substituting eqns \eqref{2240a} and \eqref{2261d}, we obtain that:
\begin{align}
F(T) =& \frac{4\,F'(0)}{b_0^2}\,\Bigg[\frac{\alpha\,\delta}{b_0\left(1+\alpha\right)}\left(2+b_0^2\,r^{-4}(T)\right)\,\left(r^2(T)+\delta\,b_0\right)\,r^2(T)+\frac{\left(1-2\alpha\right)}{\left(1+\alpha\right)}\,r^2(T)\Bigg]
\nonumber\\
&\;\times\,\exp\left[-\frac{\delta}{2b_0}\left(2\,r^2(T)-b_0^2 r^{-2}(T)\right)\right], \label{2261e}
\end{align}
where $r(T)$ is eqn \eqref{2261d}.

\item ${\bf b=3}$: Eqn \eqref{2240d} becomes:
\small
\begin{align}\label{2262d}
& 0= \frac{b_0^2\,T}{2}+b_0^2\,r^{-2}-r^{-8},
\nonumber\\
&\Rightarrow\; r^{-2}(T)=
\nonumber\\
&\; \frac{\delta_1}{2}\Bigg[\frac{b_0}{\sqrt[3]{2}\,3^{2/3}}\left(\sqrt{3}\sqrt{32T^3+27b_0^2}+9b_0\right)^{1/3}
-2b_0\sqrt[3]{\frac{2}{3}}\,T\left(\sqrt{3}\sqrt{32T^3+27b_0^2}+9b_0\right)^{-1/3}\Bigg]^{1/2}
\nonumber\\
&\;+\frac{\delta_2}{2}\Bigg[2b_0\sqrt[3]{\frac{2}{3}}\,T\left(\sqrt{3}\sqrt{32T^3+27b_0^2}+9b_0\right)^{-1/3}+2\delta_1\,b_0^2\Bigg[\frac{b_0}{\sqrt[3]{2}\,3^{2/3}}\left(\sqrt{3}\sqrt{32T^3+27b_0^2}+9b_0\right)^{1/3}
\nonumber\\
&\;-2b_0\sqrt[3]{\frac{2}{3}}\,T\left(\sqrt{3}\sqrt{32T^3+27b_0^2}+9b_0\right)^{-1/3}\Bigg]^{-1/2} -\frac{b_0}{\sqrt[3]{2}\,3^{2/3}}\left(\sqrt{3}\sqrt{32T^3+27b_0^2}+9b_0\right)^{1/3}\Bigg]^{1/2},
\end{align}
\normalsize
where $\left(\delta_1,\delta_2\right)=(\pm 1,\,\pm 1)$. By putting eqns \eqref{2240b} and \eqref{2240c} together, and then by substituting eqns \eqref{2240a} and \eqref{2262d}, we obtain that:
\begin{align}
F(T) =& \frac{4\,F'(0)}{b_0^2}\,\Bigg[\frac{\alpha\,\delta}{b_0\left(1+\alpha\right)}\left(2+b_0^2\,r^6(T)\right)\left(r^{-3}(T)+\delta\,b_0\right)\,r^{-8}(T)+\frac{\left(1+3\alpha\right)}{\left(1+\alpha\right)}\,r^{-8}(T)\Bigg]\,
\nonumber\\
&\;\times\,\exp\left[-\frac{\delta}{3b_0}\left(-2\,r^{-3}(T)+b_0^2r^3(T)\right)\right], \label{2262e}
\end{align}
where $r(T)$ is eqn \eqref{2262d}.

\end{enumerate} 


\subsubsection{$a=-\frac{1}{2}$ case solutions}\label{sect325}

\noindent For solving ${\bf a=-\frac{1}{2}}$ specific cases, eqn \eqref{2225e} will simplify as:
\begin{align}\label{2300d}
0=& \frac{b_0\,T}{2}+b_0\,r^{-2}+\delta\,r^{-2-b}.
\end{align}
Eqns \eqref{2225a} to \eqref{2225c} and \eqref{2225f} become:
\small
\begin{subequations}
\begin{align}
F'(T)=& F'(0)\,\exp\left[\int\,dr\,\frac{\left[\left(\frac{2b-1}{4}\right)\,r^{-2b}-b_0^2\right]}{\left[\frac{1}{2}\,r^{1-2b}+\delta\,b_0\,r^{1-b}\right]}\right] \label{2300a}
\\
\kappa\,\rho =& \frac{2\,F'(T)}{b_0^2\,\left(1+\alpha\right)}\,\left[-\frac{\left[\left(\frac{2b-1}{4}\right)\,r^{-2b}-b_0^2\right]}{\left[\frac{1}{2}\,r^{1-2b}+\delta\,b_0\,r^{1-b}\right]}\,\Bigg[r^{-2b-1}+\delta\,b_0\,r^{-b-1}\Bigg]+(b-\frac{1}{2})\,r^{-2b-2}\right] , \label{2300b}
\\
\kappa\,\rho =& -\frac{F(T)}{2} + \frac{2\,F'(T)}{b_0^2}\,\Bigg[-\frac{\left[\left(\frac{2b-1}{4}\right)\,r^{-2b}-b_0^2\right]}{\left[\frac{1}{2}\,r^{1-2b}+\delta\,b_0\,r^{(1-b)}\right]}\,\Bigg[r^{-2b-1}+\delta\,b_0\,r^{-b-1}\Bigg]+(b-\frac{1}{2})\,r^{-2b-2}
\nonumber\\
&\;-\frac{\delta\,b_0}{2}\,r^{-b-2}\Bigg], \label{2300c}
\\
\rho=& \rho_1\,r^{\frac{\left(1+\alpha\right)}{2\alpha}} \label{2300f}
\end{align}
\end{subequations} 
\normalsize
where $\rho_1=\rho_0\,a_0^{\frac{\left(1+\alpha\right)}{2\alpha}}=$ constant.

\noindent The possible cases are:
\begin{enumerate}
\item ${\bf b=0}$ case: Eqn \eqref{2300d} becomes:
\begin{align}\label{2301d}
0=& \frac{b_0\,T}{2}+\left(b_0+\delta\right)\,r^{-2},
\nonumber\\
&\Rightarrow\,r^{-1}(T)=\sqrt{\frac{b_0}{2\left(b_0+\delta\right)}} \sqrt{-T}.
\end{align}
By putting eqns \eqref{2300b} and \eqref{2300c} together, and then substituting eqns \eqref{2300a} and \eqref{2301d} inside, we obtain that:
\begin{align}
F(T) =& \frac{4\,F'(0)}{b_0^2}\left(\frac{b_0}{2\left(b_0+\delta\right)}\right)^{1+\frac{\left(\frac{1}{4}+b_0^2\right)}{\left(1+2\delta\,b_0\right)}}\Bigg[\frac{\alpha\left(\frac{1}{4}+b_0^2\right)\left(1+\delta\,b_0\right)}{\left(1+\alpha\right)\left(\frac{1}{2}+\delta\,b_0\right)}-\frac{\alpha}{2\left(1+\alpha\right)}-\frac{\delta\,b_0}{2}\Bigg]
\nonumber\\
&\,\times\,\left(-T\right)^{1+\frac{\left(\frac{1}{4}+b_0^2\right)}{\left(1+2\delta\,b_0\right)}},
\nonumber\\
=& F_3\,\left(-T\right)^{1+\frac{\left(\frac{1}{4}+b_0^2\right)}{\left(1+2\delta\,b_0\right)}}, \label{2301e}
\end{align}
where $F_3$ is a constant. Once again, we have a pure power-law solution as in ref \cite{baha1}.

\item ${\bf b=1}$ case: Eqn \eqref{2300d} becomes more simple as:
\begin{align}\label{2239}
0=& \frac{\delta\,b_0\,T}{2}+\delta\,b_0\,r^{-2}+r^{-3},
\nonumber\\
\;\Rightarrow\,r^{-1}(T)=&\frac{1}{3}\Bigg[-\delta\,b_0+\frac{2^{2/3}\,b_0^2}{\sqrt[3]{-4\delta\,b_0^3+3\sqrt{3}\sqrt{27b_0^2\,T^2+8b_0^4\,T}-27\delta\,b_0\,T}}
\nonumber\\
&\;+\frac{1}{2^{2/3}}\,\sqrt[3]{-4\delta\,b_0^3+3\sqrt{3}\sqrt{27b_0^2\,T^2+8b_0^4\,T}-27\delta\,b_0\,T}\Bigg]. 
\end{align}
Eqn \eqref{2239} leads to only one real solution for $r^{-1}(T)$. By putting eqns \eqref{2300b} and \eqref{2300c} together, and then substituting eqns \eqref{2300a} and \eqref{2239} inside, we obtain that:
\begin{align}
F(T)=&\frac{4\,F'(0)}{b_0^2}\,\exp\left[-\delta\,b_0\,r(T)\right]\,\Bigg[-\frac{\alpha}{\left(1+\alpha\right)}\left(\frac{r^{-1}(T)}{2}-\delta\,b_0\right)\,\left(r^{-1}(T)+\delta\,b_0\right)\,r^{-3/2}(T)
\nonumber\\
&\;+\frac{\alpha}{\left(1+\alpha\right)}\frac{r^{-7/2}(T)}{2}-\delta\,b_0\,\frac{r^{-5/2}(T)}{2}\Bigg], \label{2239e}
\end{align}
where $r(T)$ is described by eqn \eqref{2239}.

\item ${\bf b=-1}$ case: Eqn \eqref{2300d} becomes:
\begin{align}\label{2302d}
0=& \frac{b_0\,T}{2}+b_0\,r^{-2}+\delta\,r^{-1},
\nonumber\\
&\Rightarrow\; r^{-1}(T)=-\frac{\delta}{2b_0}\pm\sqrt{\frac{1}{4b_0^2}-\frac{T}{2}}.
\end{align}
By putting eqns \eqref{2300b} and \eqref{2300c} together, and then substituting eqns \eqref{2300a} and \eqref{2302d} inside, we obtain that:
\begin{align}
F(T) =& \frac{4\,F'(0)}{b_0^2}\,\Bigg[\frac{\alpha\left(\frac{3}{4}\,r^{2}(T)+b_0^2\right)\left(r(T)+\delta\,b_0\right)}{r^{2}(T)\left(1+\alpha\right)\left(\frac{r(T)}{2}+\delta\,b_0\right)}-\frac{3\alpha}{2\left(1+\alpha\right)}-\frac{\delta\,b_0}{2\,r(T)}\Bigg]\frac{\sqrt{r(T)}\,\exp\left(\frac{\delta\,b_0}{r(T)}\right)}{\left(2\delta\,b_0+r(T)\right)^2}, \label{2302e}
\end{align}
where $r(T)$ is described by eqn \eqref{2302d}.

\item ${\bf b=2}$ case: Eqn \eqref{2300d} becomes:
\begin{align}\label{2303d}
0=& \frac{\delta\,b_0\,T}{2}+\delta\,b_0\,r^{-2}+r^{-4},
\nonumber\\
&\Rightarrow\; r^{-1}(T)=\delta_2\,\sqrt{-\frac{\delta\,b_0}{2}+\delta_1\,\sqrt{\frac{b_0^2}{4}-\frac{\delta\,b_0\,T}{2}}}.
\end{align}
By putting eqns \eqref{2300b} and \eqref{2300c} together, and then substituting eqns \eqref{2300a} and \eqref{2303d} inside, we obtain that:
\begin{align}
F(T)=& \frac{4\,F'(0)}{b_0^2}\,\Bigg[-\frac{\alpha\,\left(\frac{3}{4}-b_0^2\,r^4(T)\right)\left(1+\delta\,b_0r^2(T)\right)}{\left(1+\alpha\right)\left(\frac{1}{2}+\delta\,b_0\,r^2(T)\right)}+\frac{3\alpha}{2\left(1+\alpha\right)}-\frac{\delta\,b_0r^2(T)}{2}\Bigg]
\nonumber\\
&\;\times\,\frac{r^{-\frac{9}{2}}(T)\,\exp\left(-\frac{\delta\,b_0\,r^2(T)}{2}\right)}{\sqrt{1+2\delta\,b_0\,r^2(T)}}, \label{2303e}
\end{align}
where $r(T)$ is described by eqn \eqref{2303d}.

\item ${\bf b=-2}$ case: eqn \eqref{2300d} will simplify as:
\begin{align}\label{2238}
0=& b_0+\left(\frac{b_0\,T}{2}+\delta\right)\,r^2,
\nonumber\\
&\;\Rightarrow\,r(T)=\pm \frac{\sqrt{2b_0}}{\sqrt{-b_0\,T-2\delta}},
\end{align}
where $b_0\,T+2\delta < 0$. By putting eqns \eqref{2300b} and \eqref{2300c} together, and then substituting eqns \eqref{2300a} and \eqref{2238} inside, we obtain that:
\begin{align}
F(T)=& \frac{4\,F'(0)\,\exp\left(\frac{\delta\,b_0}{2\,r^2(T)}\right)}{b_0^2\,\left(r^2(T)+2\delta\,b_0\right)^{3/2}}\,\Bigg[\frac{\alpha\,\left[\frac{5}{4}\,r^{4}(T)+b_0^2\right]\,\left[r^2(T)+\delta\,b_0\right]}{\left(1+\alpha\right)r^{3/2}(T)\,\left[\frac{1}{2}\,r^2(T)+\delta\,b_0\right]}-\frac{5\alpha\,r^{5/2}(T)}{2\left(1+\alpha\right)}
\nonumber\\
&\;-\frac{\delta\,b_0}{2}r^{1/2}(T)\Bigg], \label{2238e}
\end{align}
where $r(T)$ is the eqn \eqref{2238}.

\item ${\bf b=-3}$ case: Eqn \eqref{2300d} becomes:
\begin{align}\label{2304d}
0=& r^3+\frac{\delta b_0\,T}{2}\,r^2+\delta b_0,
\nonumber\\
\Rightarrow\; r(T)=&\frac{1}{6}\Bigg[\left[-108\delta\,b_0-\delta\,b_0^3\,T^3+6^{3/2}\,b_0\sqrt{54+b_0^2\,T^3}\right]^{1/3}
\nonumber\\
&\;+b_0^2\,T^2\left[-108\delta\,b_0-\delta\,b_0^3\,T^3+6^{3/2}\,b_0\sqrt{54+b_0^2\,T^3}\right]^{-1/3}-\delta\,b_0\,T\Bigg]  .
\end{align}
By putting eqns \eqref{2300b} and \eqref{2300c} together, and then substituting eqns \eqref{2300a} and \eqref{2304d} inside, we obtain that:
\begin{align}
F(T) =& \frac{4\,F'(0)}{b_0^2}\,\Bigg[\frac{\alpha\,\left(\frac{7}{4}\,r^6(T)+b_0^2\right)\left(r^3(T)+\delta\,b_0\right)}{\left(1+\alpha\right)r^3(T)\,\left(\frac{r^3(T)}{2}+\delta\,b_0\right)}-\frac{7\alpha\,r^3(T)}{2\left(1+\alpha\right)}-\frac{\delta\,b_0}{2}\Bigg]\frac{r^{\frac{3}{2}}(T)\,\exp\left(\frac{\delta\,b_0}{3\,r^3(T)}\right)}{\left(2\delta\,b_0+r^3(T)\right)^{\frac{4}{3}}}, \label{2304e}
\end{align}
where $r(T)$ is described by eqn \eqref{2304d}.

\item ${\bf b=4}$ case: Eqn \eqref{2300d} becomes:
\begin{align}\label{2305d}
0=& \frac{b_0\,T}{2}+b_0\,r^{-2}+\delta\,r^{-6},
\nonumber\\
\Rightarrow\; r^{-1}(T)=&\delta_1\Bigg[\frac{1}{6^{\frac{2}{3}}}\,\left[\sqrt{3}\,b_0\sqrt{16\delta b_0+27\,T^2}-9\delta\,b_0\,T\right]^{\frac{1}{3}}
\nonumber\\
&\;-\frac{\delta\,b_0\,2^{\frac{2}{3}}}{3^{\frac{1}{3}}}\left[\sqrt{3}\,b_0\sqrt{16\delta b_0+27\,T^2}-9\delta\,b_0\,T\right]^{-\frac{1}{3}}\Bigg]^{\frac{1}{2}},
\end{align}
where $\delta_1=\pm 1$. By putting eqns \eqref{2300b} and \eqref{2300c} together, and then substituting eqns \eqref{2300a} and \eqref{2305d} inside, we obtain that:
\begin{align}
F(T) =& \frac{4\,F'(0)}{b_0^2}\,\Bigg[-\frac{\alpha\,\left(\frac{7}{4}-b_0^2\,r^8(T)\right)\left(1+\delta\,b_0r^4(T)\right)}{\left(1+\alpha\right)r^4(T)\left(\frac{1}{2}+\delta\,b_0\,r^4(T)\right)}+\frac{7\alpha}{2\left(1+\alpha\right)\,r^4(T)}-\frac{\delta\,b_0}{2}\Bigg]
\nonumber\\
&\;\times\,\frac{r^{-\frac{5}{2}}(T)\,\exp\left(-\frac{\delta\,b_0\,r^4(T)}{4}\right)}{\left(1+2\delta\,b_0\,r^4(T)\right)^{\frac{3}{4}}} , \label{2305e}
\end{align}
where $r(T)$ is described by eqn \eqref{2305d}.

\item ${\bf b=-4}$ case: Eqn \eqref{2300d} becomes:
\begin{align}\label{2306d}
0=& r^4+\frac{\delta\,b_0\,T}{2}\,r^{2}+\delta\,b_0,
\nonumber\\
&\Rightarrow\; r(T)=\delta_2\sqrt{-\frac{\delta\,b_0\,T}{4}+\delta_1\sqrt{\frac{b_0^2\,T^2}{16}-\delta\,b_0}}.
\end{align}
By putting eqns \eqref{2300b} and \eqref{2300c} together, and then substituting eqns \eqref{2300a} and \eqref{2306d} inside, we obtain that:
\begin{align}
F(T) =& \frac{4\,F'(0)}{b_0^2}\,\Bigg[\frac{\alpha\,\left(\frac{9}{4}\,r^{8}(T)+b_0^2\right)\left(r^4(T)+\delta\,b_0\right)}{\left(1+\alpha\right)r^5(T)\left(\frac{r^4(T)}{2}+\delta\,b_0\right)}-\frac{9\alpha\,r^3(T)}{2\left(1+\alpha\right)}-\frac{\delta\,b_0}{2}\Bigg]\frac{r^{\frac{7}{2}}(T)\,\exp\left(\frac{\delta\,b_0}{4\,r^4(T)}\right)}{\left(2\delta\,b_0+r^4(T)\right)^{\frac{5}{4}}}, \label{2306e}
\end{align}
where $r(T)$ is described by eqn \eqref{2306d}.

\item ${\bf b=6}$ case: Eqn \eqref{2300d} becomes:
\begin{align}\label{2307d}
 0=& \frac{b_0\,T}{2}+b_0\,r^{-2}+\delta\,r^{-8},
\nonumber\\
\Rightarrow\;& r^{-2}(T)= \frac{\delta_1}{2}\Bigg[2\sqrt[3]{\frac{2}{3}}b_0\,T \left(9\delta\,b_0^2+\sqrt{3}\,b_0\sqrt{27b_0^2-32\delta\,b_0\,T^3}\right)^{-1/3}
\nonumber\\
&\;+\frac{\delta}{\sqrt[3]{2}\,3^{2/3}}\left(9\delta\,b_0^2+\sqrt{3}\,b_0\sqrt{27b_0^2-32\delta\,b_0\,T^3}\right)^{1/3}\Bigg]^{1/2}
\nonumber\\
&\;+\frac{\delta_2}{2} \Bigg[-2\sqrt[3]{\frac{2}{3}}b_0\,T\left(9\delta\,b_0^2+\sqrt{3}\,b_0\sqrt{27b_0^2-32\delta\,b_0\,T^3}\right)^{-1/3}\nonumber\\
&\;-\frac{\delta\,\delta_2}{2b_0}\Bigg[2\sqrt[3]{\frac{2}{3}}b_0\,T \left(9\delta\,b_0^2+\sqrt{3}\,b_0\sqrt{27b_0^2-32\delta\,b_0\,T^3}\right)^{-1/3}
\nonumber\\
&\;+\frac{\delta}{\sqrt[3]{2}\,3^{2/3}}\left(9\delta\,b_0^2+\sqrt{3}\,b_0\sqrt{27b_0^2-32\delta\,b_0\,T^3}\right)^{1/3}\Bigg]^{-1/2}
\nonumber\\
&\;-\frac{\delta}{\sqrt[3]{2}3^{2/3}}\left(9\delta\,b_0^2+\sqrt{3}\,b_0\sqrt{27b_0^2-32\delta\,b_0\,T^3}\right)^{1/3}\Bigg]^{1/2},
\end{align}
where the possible solutions are $\left(\delta_1,\,\delta_2\right)=\left(\pm 1,\,\pm 1 \right)$. By putting eqns \eqref{2300b} and \eqref{2300c} together, and then substituting eqns \eqref{2300a} and \eqref{2307d} inside, we obtain that:
\small
\begin{align}
F(T)= & \frac{4\,F'(0)}{b_0^2}\,\Bigg[-\frac{\alpha\,\left(\frac{11}{4}\,r^{-12}(T)-b_0^2\right)}{\left(1+\alpha\right)\left(\frac{1}{2}\,r^{-6}(T)+\delta\,b_0\right)}\,\left(r^{-6}(T)+\delta\,b_0\right)+\frac{11\alpha}{2\left(1+\alpha\right)}\,r^{-12}(T)-\frac{\delta\,b_0}{2}\,r^{-6}(T)\Bigg]
\nonumber\\
&\;\times \,\frac{r^{\frac{7}{2}}(T)}{\left(2\delta\,b_0\,r^6(T)+1\right)^{\frac{5}{6}}}\,\exp\left(-\frac{\delta\,b_0}{6}\,r^6(T)\right), \label{2307e}
\end{align}
\normalsize
where $r(T)$ is described by eqn \eqref{2307d}.

\item ${\bf b=-6}$ case: Eqn \eqref{2300d} becomes:
\begin{align}\label{2308d}
0=& r^6+\frac{\delta\,b_0\,T}{2}\,r^{2}+\delta\,b_0,
\nonumber\\
&\Rightarrow\; r^2(T)=\frac{\left[\sqrt{6}\,b_0\sqrt{\delta\,b_0\,T^3+54}-18\delta\,b_0\right]^{2/3}-\sqrt[3]{6}\,\delta\,b_0\,T}{6^{2/3}\left[\sqrt{6}\,b_0\sqrt{\delta\,b_0\,T^3+54}-18\delta\,b_0\right]^{1/3}}
\end{align}
By putting eqns \eqref{2300b} and \eqref{2300c} together, and then substituting eqns \eqref{2300a} and \eqref{2308d} inside, we obtain that:
\small
\begin{align}
F(T) =& \frac{4\,F'(0)}{b_0^2}\,\Bigg[\frac{\alpha\,\left(\frac{13}{4}\,r^{12}(T)+b_0^2\right)}{r^6(T)\left(1+\alpha\right)\left(\frac{1}{2}\,r^{6}(T)+\delta\,b_0\right)}\,\left(r^{6}(T)+\delta\,b_0\right)-\frac{13\alpha}{2\left(1+\alpha\right)}\,r^{6}(T)-\frac{\delta\,b_0}{2}\Bigg]
\nonumber\\
&\,\times\,\exp\left(\frac{\delta\,b_0}{6\,r^6(T)}\right)\,\frac{r^{9/2}(T)}{\left(2\delta\,b_0+r^6(T)\right)^{\frac{7}{6}}}, \label{2308e}
\end{align}
\normalsize
where $r(T)$ is described by eqn \eqref{2308d}.

\item ${\bf b=-8}$ case: Eqn \eqref{2300d} becomes:
\begin{align}\label{2309d}
0=& r^8+\frac{\delta\,b_0\,T}{2}\,r^{2}+\delta\,b_0,
\nonumber\\
\Rightarrow &\; r^2(T)= \frac{\delta_1}{2}\Bigg[\frac{8\delta\,b_0}{\sqrt[3]{3}}\left(\sqrt{3}b_0\sqrt{27b_0^2\,T^4-4096\delta\,b_0}+9b_0^2\,T^2\right)^{-1/3}
\nonumber\\
&\;+\frac{1}{2\cdot\,3^{2/3}}\left(\sqrt{3}b_0\sqrt{27b_0^2\,T^4-4096\delta\,b_0}+9b_0^2\,T^2\right)^{1/3}\Bigg]^{1/2}
\nonumber\\
&\;+\frac{\delta_2}{2}\Bigg[-\frac{8\delta\,b_0}{\sqrt[3]{3}}\left(\sqrt{3}b_0\sqrt{27b_0^2\,T^4-4096\delta\,b_0}+9b_0^2\,T^2\right)^{-1/3}
\nonumber\\
&\;-\delta_1\delta\,b_0\,T\Bigg[\frac{8\delta\,b_0}{\sqrt[3]{3}}\left(\sqrt{3}b_0\sqrt{27b_0^2\,T^4-4096\delta\,b_0}+9b_0^2\,T^2\right)^{-1/3}
\nonumber\\
&\;+\frac{1}{2\cdot\,3^{2/3}}\left(\sqrt{3}b_0\sqrt{27b_0^2\,T^4-4096\delta\,b_0}+9b_0^2\,T^2\right)^{1/3}\Bigg]^{-1/2}
\nonumber\\
&\;-\frac{1}{2\cdot\,3^{2/3}}\left(\sqrt{3}b_0\sqrt{27b_0^2\,T^4-4096\delta\,b_0}+9b_0^2\,T^2\right)^{1/3}\Bigg]^{1/2}
\end{align}
By putting eqns \eqref{2300b} and \eqref{2300c} together, and then substituting eqns \eqref{2300a} and \eqref{2309d} inside, we obtain that:
\small
\begin{align}
F(T) =&  \frac{4\,F'(0)}{b_0^2}\,\Bigg[\frac{\alpha\,\left(\frac{17}{4}\,r^{16}(T)+b_0^2\right)}{r^{8}(T)\left(1+\alpha\right)\left(\frac{1}{2}\,r^{8}(T)+\delta\,b_0\right)}\,\left(r^{8}(T)+\delta\,b_0\right)-\frac{17\alpha}{2\left(1+\alpha\right)}\,r^{8}(T)-\frac{\delta\,b_0}{2}\Bigg]
\nonumber\\
&\,\times\,\exp\left(\frac{\delta\,b_0}{8\,r^8(T)}\right)\,\frac{r^{13/2}(T)}{\left(2\delta\,b_0+r^8(T)\right)^{\frac{9}{8}}}, \label{2309e}
\end{align}
\normalsize
where $r(T)$ is described by eqn \eqref{2309d}.

\end{enumerate}

In this section, all these previous non power-law teleparallel $F(T)$ solutions are new. We may also use several different coframe ansatz leading to additional new $F(T)$ solutions. Eqn \eqref{2160a} power-law ansatz based $F(T)$ solutions are sufficient for the current paper's aims and purposes. We may study several specific cases such as radiation fluids $\alpha=\frac{1}{3}$ to name this example \cite{hawkingellis1,coleybook}. We are also able to study physical properties of possible singularities arising from each new previous $F(T)$ solutions. Even if there are numerous new and more complex singularities in these previous $F(T)$ solutions, they may lead to some possible black hole solutions (point-like or not) and/or matter absorbing points. This task is beyond the aims of the paper and might be for potential future works.


\section{Dust perfect fluid solutions ($\alpha=0$)}\label{sect33}

This specific case arises from $P(r)=0$ and $\rho(r) \neq 0$ consideration. By setting $\alpha=0$ inside eqn \eqref{2210}, the conservation law becomes:
\begin{align}
A_1'(r)=0 .  \label{2501}
\end{align}
We need that $A_1(r)=a_0=$ constant. Then eqn \eqref{2201a} remains unchanged, but eqns \eqref{2212b} and \eqref{2212c} will be simplified:
\begin{subequations}
\begin{align}
\kappa\,\rho =& 2\,F'(T)\,\left[-\left(\frac{g_1(r)}{k_1(r)}\right)\,k_2(r)+g_2(r)\right] , \label{2501a}
\\
\kappa\,\rho =& -\frac{F(T)}{2} + 2\,F'(T)\,\left[-\left(\frac{g_1(r)}{k_1(r)}\right)\,k_2(r)+g_3(r)\right] . \label{2501b}
\end{align}
\end{subequations} 
By combining eqns \eqref{2501a} and \eqref{2501b} and substituting eqns \eqref{2922b} and \eqref{2922c} FE components, we will obtain a simplified relation for $F(T(r))$ as:
\small
\begin{align}
F(T(r))=& 4\,F'(T(r))\,\left[g_3(r)-g_2(r)\right] ,
\nonumber\\
=& -\frac{4\,F'(T(r))\,A_3'}{A_2\,A_3^2}\left(A_3'+\delta\,A_2\right) . \label{2502}
\end{align}
\normalsize
Eqn \eqref{2201a} becomes:
\begin{align}
F'(T)=& F'(0)\exp\Bigg[\int\,dr\,\frac{\left[-A_2\,A_3\,A_3''+A_2\,A_3'^2+A_2'\,A_3\,A_3'-A_2^3\right]}{A_2\,A_3\,\left(A_3'+\delta\,A_2\right)}\Bigg] , \label{2501c}
\end{align}
Eqn \eqref{2213b} for torsion scalar becomes:
\begin{align}\label{2501d}
T(r) = -2\left(\frac{\delta}{A_3}+\frac{A_3'}{A_2\,A_3}\right)^2.
\end{align}
As for previous cases, we will apply the $A_3=r$ coordinate set. The $A_3=c_0=$ constant coordinate leads to constant torsion scalar and GR solutions, which is not relevant for the current purpose.

\noindent For $A_3=r$ coordinate system, eqns \eqref{2502} to \eqref{2501d} become:
\begin{subequations}
\begin{align}
F(T)=& -\frac{4\,F'(T)}{A_2^2\,r^2}\left(1+\delta\,A_2\right) , \label{2504a}
\\
F'(T)=& F'(0)\,\frac{r\,A_2}{\left(1+\delta\,A_2\right)}\exp\Bigg[-\delta\,\int\,dr\,\frac{A_2}{r}\Bigg]  , \label{2504b}
\\
T=& T(r) = -\frac{2}{A_2^2\,r^2}\left(1+\delta\,A_2\right)^2 \label{2504c}.
\end{align}
\end{subequations}
By substituting eqn \eqref{2504b} into eqn \eqref{2504a}, we find that:
\begin{align}
F(T)=& -\frac{4\,F'(0)}{r\,A_2}\,\exp\Bigg[-\delta\,\int\,dr\,\frac{A_2}{r}\Bigg] , \label{2505}
\end{align}
The best way for solving eqns \eqref{2504b} to \eqref{2505} is by a power-law solution ansatz as $A_2(r)=b_0\,r^b$. Note also that by setting $A_2(r)=\left(1-k\,r^2\right)^{-1/2}$ for static Robertson-Walker spacetimes, we obtain that $F(T)$ will be linear and this is a GR solution. The eqns \eqref{2504b} to \eqref{2505} become:
\begin{subequations}
\begin{align}
F(T)=& -\frac{4\,F'(0)}{b_0\,r^{b+1}}\,\exp\Bigg[-\frac{\delta\,b_0}{b}\,r^b\Bigg] , \label{2506a}
\\
F'(T)=& \frac{F'(0)\,b_0\,r^{b+1}}{\left(1+\delta\,b_0\,r^b\right)}\exp\Bigg[-\frac{\delta\,b_0}{b}\,r^b\Bigg]= -\frac{b_0^2\,r^{2(b+1)}}{4\left(1+\delta\,b_0\,r^b\right)}\,F(T), \label{2506b}
\\
T(r) =& -\frac{2}{b_0^2\,r^{2(b+1)}}\left(1+\delta\,b_0\,r^b\right)^2 \label{2506c},
\end{align}
\end{subequations}
where $b\neq 0$. The case $b=0$ is the simple static cosmological spacetime and this case may be considered as a special case. By substituting eqns \eqref{2506a} and \eqref{2506c} into eqn \eqref{2506b}, we obtain the simplified DE to solve for $F(T)$ in a cosmological dust fluid where $b\neq 0$:
\begin{align}\label{2506}
T\,F'(T) =& \frac{\left(1+\delta\,b_0\,r^b(T)\right)}{2}\,F(T).
\end{align}
By using eqn \eqref{2501a} and then substituting eqns \eqref{2506} and \eqref{2923i}, we find the fluid density:
\begin{align}\label{2506density}
\rho(T) =& \frac{F(T)}{\kappa\,b_0^2\,(-T)}\,r^{-2b-2}(T)\,\left(1+\delta\,b_0\,r^b(T)\right)^2\left(1-\delta\,b_0\,r^b(T)\right),
\end{align}
where $F(T)$ is given by eqn \eqref{2506} solutions. We will solve this eqn \eqref{2506} for some values of $b$. For pure $F(T)$ solutions, we need to find from eqn \eqref{2506c} the characteristic  equation and then solve for $r(T)$:
\begin{align}\label{2507}
0=r^{b+1}-\sqrt{-\frac{2}{T}}\,r^b-\frac{\delta}{b_0}\,\sqrt{-\frac{2}{T}}.
\end{align}
There are some specific value of $b$ leading to analytic $r(T)$ function and then to $F(T)$ solution:
\begin{enumerate}
\item ${\bf b=0}$: For this simple case of cosmological spacetime, eqn \eqref{2507} becomes:
\begin{align}\label{2381d}
r^2(T)=-\left(1+\frac{\delta}{b_0}\right)^2\,\frac{2}{T}.
\end{align}
The eqns \eqref{2504a} to \eqref{2504c} for $A_2=b_0$ will be summarized by the same eqn \eqref{2506}:
\begin{align}\label{2381a}
T\,F'(T) = \frac{\left(1+\delta\,b_0\right)}{2}\,F(T)
\end{align}
We solve this eqn \eqref{2381a} and obtain as solution for a flat dust fluid:
\begin{align}
F(T)=& F_0\,T^{\frac{1+\delta\,b_0}{2}}, \label{2381e}
\end{align}
where $b_0 \neq \pm\delta$ for a teleparallel solution (i.e. $b_0 = \delta$ leads to GR solutions). Once again, we obtain a pure power-law solution as in ref. \cite{SSpaper} for general $X_4$ similarity (here $\rho=\rho(r)$ without any other constraint). By using eqn \eqref{2506density}, setting $b=0$ and substituting eqn \eqref{2381e}, the fluid density $\rho(T)$ is:
\begin{align}\label{2381ea}
\rho(T) =& \frac{F_0}{2\kappa}\,\left(1-\delta\,b_0\right)\,T^{\frac{\delta\,b_0+1}{2}} .
\end{align}
Eqn \eqref{2381ea} is again a power-law function of $T$ as usual. If $b_0=\delta$, we find that $\rho(T)=0$ for $F(T)=F_0\,T$.

\item ${\bf b=1}$: Eqn \eqref{2507} becomes:
\begin{align}\label{2507a}
&0=r^{2}-\sqrt{-\frac{2}{T}}\,r-\frac{\delta}{b_0}\,\sqrt{-\frac{2}{T}},
\nonumber\\
& \Rightarrow\;r(T)=\frac{1}{\sqrt{(-2T)}}\left[1+\delta_1 \sqrt{1+\frac{2\delta}{b_0}\sqrt{(-2T)}}\right],
\end{align}
where $\delta_1=\pm 1$. Then eqns \eqref{2506} becomes:
\begin{align}
T\,F'(T) =\frac{F(T)}{2} \left[1+\frac{\delta\,b_0}{\sqrt{(-2T)}}\left[1+\delta_1 \sqrt{1+\frac{2\delta}{b_0}\sqrt{(-2T)}}\right]\right] . \label{2506ab}
\end{align}
The solution of eqn \eqref{2506ab} is:
\begin{align}
F(T)=& F_0\,\sqrt{-T}\,\left[\frac{1-\sqrt{1+\frac{2\sqrt{2}\delta}{b_0}\sqrt{-T}}}{1+\sqrt{1+\frac{2\sqrt{2}\delta}{b_0}\sqrt{-T}}}\right]^{\delta_1}\,\exp\left[-\frac{\delta\,b_0}{\sqrt{2}\,\sqrt{-T}}\left(1+\delta_1\sqrt{1+\frac{2\delta}{b_0}\sqrt{(-2T)}}\right)\right], \label{2506ae}
\end{align}
where $T\leq 0$. The fluid density $\rho(T)$ will be from eqn \eqref{2506density}:
\begin{align}
\rho(T) =& \frac{4F_0\,(-T)^{3/2}}{b_0^2\kappa\,\left[1+\delta_1 \sqrt{1+\frac{2\delta}{b_0}\sqrt{(-2T)}}\right]^4}\,\left[1+\frac{\delta\,b_0}{\sqrt{(-2T)}}\left[1+\delta_1 \sqrt{1+\frac{2\delta}{b_0}\sqrt{(-2T)}}\right]\right]^2
\nonumber\\
&\;\times\,\left[1-\frac{\delta\,b_0}{\sqrt{(-2T)}}\left[1+\delta_1 \sqrt{1+\frac{2\delta}{b_0}\sqrt{(-2T)}}\right]\right]\,\left[\frac{1-\sqrt{1+\frac{2\sqrt{2}\delta}{b_0}\sqrt{-T}}}{1+\sqrt{1+\frac{2\sqrt{2}\delta}{b_0}\sqrt{-T}}}\right]^{\delta_1}
\nonumber\\
&\;\times\,\exp\left[-\frac{\delta\,b_0}{\sqrt{2}\,\sqrt{-T}}\left[1+\delta_1\sqrt{1+\frac{2\delta}{b_0}\sqrt{(-2T)}}\right]\right].
\end{align}

\item ${\bf b=-1}$: Eqn \eqref{2507} becomes:
\begin{align}\label{2507b}
& 0=1-\sqrt{-\frac{2}{T}}\,r^{-1}-\frac{\delta}{b_0}\,\sqrt{-\frac{2}{T}}.
\nonumber\\
& \Rightarrow\;r^{-1}(T)=\sqrt{-\frac{T}{2}}-\frac{\delta}{b_0}.
\end{align}
Eqns \eqref{2506} becomes a simple DE:
\begin{align}
\frac{dF}{d\left(-T\right)}=& \frac{\delta b_0}{2\sqrt{2}}\,\left(-T\right)^{-1/2}\,F(T)  . \label{2506bd}
\end{align}
The solution of eqn \eqref{2506bd} is:
\begin{align}
F(T)=& F_1\,\exp\left[\frac{\delta b_0}{\sqrt{2}}\,\sqrt{-T}\right]  , \label{2506be}
\end{align}
where $T \leq 0$ and $F_1$ is an integration constant. The fluid density $\rho(T)$ will be from eqn \eqref{2506density}:
\begin{align}
\rho(T) =& \frac{F_1}{\kappa}\,\left[1-\frac{\delta b_0\sqrt{-T}}{2\sqrt{2}}\right]\exp\left[\frac{\delta b_0}{\sqrt{2}}\,\sqrt{-T}\right].
\end{align}

\item ${\bf b=-2}$: Eqn \eqref{2507} becomes:
\begin{align}\label{2507d}
& 0=r^{-1}-\sqrt{-\frac{2}{T}}\,r^{-2}-\frac{\delta}{b_0}\,\sqrt{-\frac{2}{T}},
\nonumber\\
& \Rightarrow\;r^{-1}(T)= \sqrt{-\frac{T}{8}}+\delta_1\sqrt{-\frac{T}{8}-\frac{\delta}{b_0}} ,
\end{align}
where $T \leq 0$ and $\delta_1=\pm 1$. Eqn \eqref{2506} will be a DE and the solution is:
\small
\begin{align}
F(T)=& F_2\,\exp\left[-\frac{\delta\,b_0}{8}\,T\,\left(1+\delta_1\,\sqrt{1+\frac{8\delta}{b_0\,T}}\right)\right] \,\left[\frac{1-\sqrt{1+\frac{8\delta}{b_0\,T}}}{1+\sqrt{1+\frac{8\delta}{b_0\,T}}}\right]^{\frac{\delta_1}{2}}
,  \label{2506de}
\end{align}
\normalsize
where $F_2$ is an integration constant. The fluid density $\rho(T)$ will be from eqn \eqref{2506density}:
\small
\begin{align}
\rho(T) =& \frac{8F_2}{\kappa\,b_0^2\,T^2}\left(1+\delta_1\sqrt{1+\frac{8\delta}{b_0\,T}}\right)^{-2} \left[1-\frac{\delta\,b_0}{8}\,T\left(1+\delta_1\sqrt{1+\frac{8\delta}{b_0\,T}}\right)^2\right]^2\,\left[\frac{1-\sqrt{1+\frac{8\delta}{b_0\,T}}}{1+\sqrt{1+\frac{8\delta}{b_0\,T}}}\right]^{\frac{\delta_1}{2}}
\nonumber\\
&\;\times\,\left[1+\frac{\delta\,b_0}{8}\,T\left(1+\delta_1\sqrt{1+\frac{8\delta}{b_0\,T}}\right)^2\right]
\exp\left[-\frac{\delta\,b_0}{8}\,T\,\left(1+\delta_1\,\sqrt{1+\frac{8\delta}{b_0\,T}}\right)\right] .
\end{align}
\normalsize

\item ${\bf b=\left\lbrace 2,\,3,\,-3,\,-4 \right\rbrace}$: We can in principle find analytic $r(T)$ solutions to the eqn \eqref{2507} characteristic equation. However these $r(T)$ cannot lead to solvable and well-defined $F(T)$ solutions and this is explaining the limited number of possible power-law ansatz analytical $F(T)$ solutions for dust fluids.

\end{enumerate}

All these teleparallel $F(T)$ solutions found in this section are all new. We may also use several other possible ansatz for finding further new $F(T)$ solutions as for section \ref{sect3}. However we only used in this section power-law ansatz as defined by eqn \eqref{2160a} with $a=0$ (because eqn \eqref{2501}) and solved for several new and interesting $F(T)$ solutions all useful for many types of astrophysical or cosmological dust fluids. We may still study and look in detals for singularity and related physical characteristics of them in potential future works as for section \ref{sect3} solutions. We can also find some point-like singularity solutions and/or matter absorbing singularities in these new $F(T)$ solutions.


\section{Non-linear perfect fluid solutions}\label{sect4}

Another class of non-vacuum solutions assumes an EoS $P(r)=\alpha\,\rho(r)+\beta\,\rho^w(r)$ with $-1 < \alpha \leq 1$, $w >1$ where it is often assumed that $\beta\,\rho^{w-1}(r) \ll \alpha$. The second term of this EoS can describe non-linear dissipating terms. Eqn \eqref{2201d} will simplify as follows:
\begin{align}
\left(\ln\,A_1\right)'+\frac{\left[\alpha+\beta\,w\,\rho^{w-1}\right]}{\left[\left(1+\alpha\right)\rho +\beta\,\rho^w\right]}\rho'=0.  \label{2251}
\end{align}
The general solution is:
\begin{align}\label{2252}
A_1(r)=A_1(0)\,\left[\left(1+\alpha\right)\rho^{1-w} +\beta\right]^{\left[\frac{\alpha}{\left(1+\alpha\right)\left(w-1\right)}-\frac{w}{w-1}\right]}\,\rho^{-w}
\end{align}
We need to set the FEs for power-$w$ fluid density EoS and then solve for new $F(T)$ solutions. We will have that eqns \eqref{2201a} and \eqref{2213b} remain unchanged and then eqns \eqref{2202b} and \eqref{2202c} will be as:
\begin{subequations}
\begin{align}
\left(1+\alpha\right)\,\left(\kappa\rho\right)+\kappa^{1-w}\,\beta\,\left(\kappa\rho\right)^w =& 2\,F'(T)\,\left[-\frac{g_1(r)}{k_1(r)}\,k_2(r)+g_2(r)\right] , \label{2253b}
\\
\kappa\,\rho =& -\frac{F(T)}{2} + 2\,F'(T)\,\left[-\frac{g_1(r)}{k_1(r)}\,k_2(r)+g_3(r)\right] . \label{2253c}
\end{align}
\end{subequations} 
By putting eqns \eqref{2253b} and \eqref{2253c} together, we find the unified equation linking $F(T)$ and $F'(T)$:
\begin{align}
&\left(1+\alpha\right)\,\Bigg[-\frac{F(T)}{2} + 2\,F'(T)\,\left[-\frac{g_1(r)}{k_1(r)}\,k_2(r)+g_3(r)\right]\Bigg]
\nonumber\\
&\;+\kappa^{1-w}\,\beta\,\Bigg[-\frac{F(T)}{2} + 2\,F'(T)\,\left[-\frac{g_1(r)}{k_1(r)}\,k_2(r)+g_3(r)\right]\Bigg]^w = 2\,F'(T)\,\left[-\frac{g_1(r)}{k_1(r)}\,k_2(r)+g_2(r)\right].
 \label{2254}
\end{align}
There are in principle several possible ansatz for solving the system governed by eqns \eqref{2201a}, \eqref{2213b}, \eqref{2253c} and \eqref{2254}, completed by eqn \eqref{2252} conservation laws solution. We will present some possible solvable solutions.


\subsection{$A_3=$ constant power-law solutions}\label{sect42}

By using the eqn \eqref{2160a} ansatz and setting $A_3=c_0=$ constant, we will solve eqns \eqref{2202a}, \eqref{2230a}, \eqref{2252}. \eqref{2253b} and \eqref{2253c}: 
\small
\begin{subequations}
\begin{align}
& F'(T)= F'(0)\,\exp\left[\int\,dr\,\frac{\left[\left(a(1-a+b)\right)\,r^{-2b-2}-\left(\frac{b_0}{c_0}\right)^2\right]}{\left[a\,r^{-2b-1}+\delta\,\left(\frac{b_0}{c_0}\right)\,r^{-b}\right]}\right] , \label{2570a}
\\
&\left(1+\alpha\right)\,\left(\kappa\rho\right)+\kappa^{1-w}\,\beta\,\left(\kappa\rho\right)^w = -\frac{2\delta}{b_0\,c_0}\,F'(T)\,\Bigg[\frac{\left[\left(a(1-a+b)\right)\,r^{-2b-2}-\left(\frac{b_0}{c_0}\right)^2\right]}{\left[a\,r^{-b-1}+\delta\,\left(\frac{b_0}{c_0}\right)\right]}\Bigg] , \label{2570b}
\\
&\kappa\,\rho = -\frac{F(T)}{2} - \frac{2\delta}{b_0\,c_0} \,F'(T)\,\Bigg[\frac{\left[\left(a(1-a+b)\right)\,r^{-2b-2}-\left(\frac{b_0}{c_0}\right)^2\right]}{\left[a\,r^{-b-1}+\delta\,\left(\frac{b_0}{c_0}\right)\right]}+a\,r^{-b-1}\Bigg] , \label{2570c}
\\
&T(r) = -\frac{2}{c_0^2}-\frac{4\delta a}{b_0\,c_0}\,r^{-(b+1)},  \label{2570d}
\\
&\tilde{a}_0\,r^a= \left[\left(1+\alpha\right)\rho^{1-w} +\beta\right]^{\left[\frac{\alpha}{\left(1+\alpha\right)\left(w-1\right)}-\frac{w}{w-1}\right]}\,\rho^{-w} , \label{2570f}
\end{align}
\end{subequations}
\normalsize
where $\tilde{a}_0$ is a constant from conservation laws. From eqn \eqref{2570d}, $a= 0$ and/or $b= -1$ leads to GR solutions (because constant torsion scalar). For all other cases, we isolate $r(T)$ from this eqn \eqref{2570d}:
\begin{align}\label{2570e}
r(T)=& \left(-\frac{4\delta a}{b_0\,c_0}\right)^{\frac{1}{b+1}}\left(T+\frac{2}{c_0^2}\right)^{-\frac{1}{b+1}} , 
\nonumber\\
dr=& -\frac{r(T)}{(b+1)}\,\frac{dT}{\left(T+\frac{2}{c_0^2}\right)},
\end{align}
where $b\neq -1$. Then eqns \eqref{2570a} to \eqref{2570c} become:
\begin{subequations}
\begin{align}
& F'(T)=F'(0)\,\left(-c_0^2\right)^{1-\frac{a}{(b+1)}} \left(2-c_0^2\,T\right)^{\frac{2a}{(b+1)}-1}\left(2+c_0^2\,T\right)^{-\frac{a}{(b+1)}}\exp\left[\frac{4a}{(b+1)\left(2+c_0^2\,T\right)}\right]
 \label{2571a}
\\
&\left(1+\alpha\right)\,\left(\kappa\rho\right)+\kappa^{1-w}\,\beta\,\left(\kappa\rho\right)^w = \frac{F'(T)}{2}\,\Bigg[\frac{\left(\frac{b+1}{a}-1\right)\left(T+\frac{2}{c_0^2}\right)^{2}-\frac{16}{c_0^4}}{\left(T-\frac{2}{c_0^2}\right)}\Bigg] , \label{2571b}
\\
&\kappa\,\rho = -\frac{F(T)}{2} + \frac{F'(T)}{2}\,\Bigg[\frac{\left(\frac{b+1}{a}-1\right)\left(T+\frac{2}{c_0^2}\right)^{2}-\frac{16}{c_0^4}}{\left(T-\frac{2}{c_0^2}\right)}+\left(T+\frac{2}{c_0^2}\right)\Bigg] . \label{2571c}
\end{align}
\end{subequations}
By putting eqns \eqref{2571b} and \eqref{2571c} together, we obtain that:
\small
\begin{align}
&\Bigg[-F(T)+F'(T)\left(T+\frac{2}{c_0^2}\right)\Bigg]+\alpha\,\Bigg[-F(T) + F'(T)\,\Bigg[\frac{\left(\frac{b+1}{a}-1\right)\left(T+\frac{2}{c_0^2}\right)^{2}-\frac{16}{c_0^4}}{\left(T-\frac{2}{c_0^2}\right)}+\left(T+\frac{2}{c_0^2}\right)\Bigg]\Bigg]
\nonumber\\
&\;+\frac{\beta}{(2\kappa)^{w-1}}\,\Bigg[-F(T) + F'(T)\,\Bigg[\frac{\left(\frac{b+1}{a}-1\right)\left(T+\frac{2}{c_0^2}\right)^{2}-\frac{16}{c_0^4}}{\left(T-\frac{2}{c_0^2}\right)}+\left(T+\frac{2}{c_0^2}\right)\Bigg]\Bigg]^w = 0 , \label{2571d}
\end{align}
\normalsize
where $F(T) \neq F'(T)\left(T+\frac{2}{c_0^2}\right)$. Otherwise, we obtain a linear $F(T)$ leading to GR solutions. Eqn \eqref{2571d} can also be written as the form:
\small
\begin{align}\label{2571e}
\Bigg[-F(T)+F'(T)\left(T+\frac{2}{c_0^2}\right)\Bigg]+\alpha\,G_1\left(T,\,F(T),\,F'(T)\right)+\frac{\beta}{(2\kappa)^{w-1}}\,\left[G_1\left(T,\,F(T),\,F'(T)\right)\right]^w = 0, 
\end{align}
\normalsize
where the function $G_1\left(T,\,F(T),\,F'(T)\right)$ is
\begin{align}\label{2571f}
G_1\left(T,\,F(T),\,F'(T)\right)=-F(T) + F'(T)\,\Bigg[\frac{\left(\frac{b+1}{a}-1\right)\left(T+\frac{2}{c_0^2}\right)^{2}-\frac{16}{c_0^4}}{\left(T-\frac{2}{c_0^2}\right)}+\left(T+\frac{2}{c_0^2}\right)\Bigg]. 
\end{align}
We may solve eqn \eqref{2571e} describing a polynomial of degree $w$ only for $w=2$, $3$ and $4$, because $G_1$ is linear in $F(T)$ and $F'(T)$. For $w=2$ ,we obtain as solution to eqn \eqref{2571e}:
\small
\begin{align}\label{2571g}
&F(T) - F'(T) \Bigg[\frac{\left(\frac{b+1}{a}-1\right)\left(T+\frac{2}{c_0^2}\right)^{2}-\frac{16}{c_0^4}}{\left(T-\frac{2}{c_0^2}\right)}+\left(T+\frac{2}{c_0^2}\right)\Bigg]
\nonumber\\
&\quad=\frac{\alpha\,\kappa}{\beta}\left[1-\delta_2\sqrt{1+\frac{2\beta}{\alpha^2\kappa}\left(F(T)-F'(T)\left(T+\frac{2}{c_0^2}\right)\right)}\right]
\end{align}
\normalsize
The eqn \eqref{2571g} is difficult to solve for an exact solution because the square root to the r.h.s. But if we use the approximation $\beta \ll \alpha$ for slightly non-linear fluid approximation, then eqn \eqref{2571g} will simplify and becomes a linear DE:
\begin{align}\label{2571h}
\frac{\left(\alpha+\delta_2\right)\left(T-\frac{2}{c_0^2}\right)}{\Bigg[\left(\alpha\,\left(\frac{b+1}{a}\right)+\delta_2\right)T^2+\frac{4\alpha}{c_0^2}\left(\frac{b+1}{a}-1\right)T+\frac{4}{c_0^4}\left(\alpha\left(\frac{b+1}{a}-6\right)-\delta_2 \right) \Bigg]} = \frac{F'(T)}{\left[ F(T)-\frac{\alpha^2\,\kappa\left(1-\delta_2\right)}{\beta\left(\alpha+\delta_2\right)}\right]}  .
\end{align}
The general solution of eqn \eqref{2571h} is:
\small
\begin{align}\label{2571i}
F(T)=&\frac{\alpha^2\,\kappa\left(1-\delta_2\right)}{\beta\left(\alpha+\delta_2\right)}+\left[ F(0)-\frac{\alpha^2\,\kappa\left(1-\delta_2\right)}{\beta\left(\alpha+\delta_2\right)}\right]
\nonumber\\
&\;\times \left[\left(4\alpha(b+1-6a)-4a\alpha_2\right) +\left(4\alpha\,c_0^2(b+1-a)\right)\,T+\left(c_0^4\,(a\,\delta_2+\alpha\,(b+1))\right)\,T^2\right]^{\frac{\left(\alpha+\delta_2\right)\,a}{2\left(\delta_2\,a+\alpha (1+b)\right)}}
\nonumber\\ 
&\;\times\exp\Bigg[\frac{\left(\alpha+\delta_2\right)\sqrt{a}\left(\alpha(a-2(b+1))-a\,\delta_2\right)}{\left(\delta_2\,a+\alpha (1+b)\right) \sqrt{6a\alpha\,\delta_2+a+\alpha^2(a+4(b+1))}}
\nonumber\\ 
&\;\times\tanh^{-1}\left(\frac{2\alpha (a-b-1)-c_0^2(a\delta_2+\alpha (b+1))\,T}{2\sqrt{a}\sqrt{6a\alpha\,\delta_2+a+\alpha^2(a+4(b+1))}}\right)\Bigg].
\end{align}
\normalsize
For $\delta_2=+1$, the solution will be the same as a linear perfect fluid which is solved in section \ref{sect3}. The most interesting case is $\delta_2=-1$ where eqn \eqref{2571i} becomes:
\small
\begin{align}\label{2571j}
F(T)=&\frac{2\alpha^2\,\kappa}{\beta\left(\alpha-1\right)}+\left[ F(0)-\frac{2\alpha^2\,\kappa}{\beta\left(\alpha-1\right)}\right]
\nonumber\\
&\;\times \left[\left(4\alpha(b+1-6a)+4a\right) +\left(4\alpha\,c_0^2(b+1-a)\right)\,T+\left(c_0^4\,(-a+\alpha\,(b+1))\right)\,T^2\right]^{\frac{\left(\alpha-1\right)\,a}{2\left(-a+\alpha (1+b)\right)}}
\nonumber\\ 
&\;\times\exp\Bigg[\frac{\left(\alpha-1\right)\sqrt{a}\left(\alpha(a-2(b+1))+a\right)}{\left(-a+\alpha (1+b)\right) \sqrt{-6a\alpha+a+\alpha^2(a+4(b+1))}}
\nonumber\\ 
&\;\times\tanh^{-1}\left(\frac{2\alpha (a-b-1)-c_0^2(-a+\alpha (b+1))\,T}{2\sqrt{a}\sqrt{-6a\alpha+a+\alpha^2(a+4(b+1))}}\right)\Bigg]
\end{align}
\normalsize
where $\alpha \neq 1$. We found at eqn \eqref{2571j} a real quadratic new $F(T)$ solution for a weak correction in $\rho^2$ to the linear and isotropic perfect fluid. We can proceed to the same exercise for $w=3$ and $w=4$ corrections, we will just obtain a slightly different $F(T)$ solution in both cases.

\subsection{$A_3=r$ power-law solutions}\label{sect43}

By using the eqn \eqref{2160a} ansatz and setting $A_3=r$ as coordinate choice, we will solve \eqref{2202a}, \eqref{2224d}, \eqref{2252}, \eqref{2253b} and \eqref{2253c}: 
\small
\begin{subequations}
\begin{align}
& F'(T)= F'(0)\,\exp\left[\int\,dr\,\frac{\left[\left(2a-a^2+ab+b+1\right)\,r^{-2b}-b_0^2\right]}{\left[(a+1)\,r^{2(1-b)-1}+\delta\,b_0\,r^{(1-b)}\right]}\right] , \label{2580a}
\\
&\left(1+\alpha\right)\,\left(\kappa\rho\right)+\kappa^{1-w}\,\beta\,\left(\kappa\rho\right)^w = \frac{2}{b_0^2}\,F'(T)\,\Bigg[-\frac{\left[\left(2a-a^2+ab+b+1\right)\,r^{-2b}-b_0^2\right]}{\left[(a+1)\,r^{2(1-b)-1}+\delta\,b_0\,r^{(1-b)}\right]}\Bigg[r^{-2b-1}+\delta\,b_0\,r^{-b-1}\Bigg]
\nonumber\\
&+(a+b)\,r^{-2b-2}\Bigg] , \label{2580b}
\\
&\kappa\,\rho = -\frac{F(T)}{2} + \frac{2}{b_0^2}\,F'(T)\,\Bigg[-\frac{\left[\left(2a-a^2+ab+b+1\right)\,r^{-2b}-b_0^2\right]}{\left[(a+1)\,r^{2(1-b)-1}+\delta\,b_0\,r^{(1-b)}\right]}\Bigg[r^{-2b-1}+\delta\,b_0\,r^{-b-1}\Bigg]
\nonumber\\
&+\Bigg[(-a+b-1)\,r^{-2b-2}-\delta\,b_0\,(a+1)\,r^{-b-2}\Bigg]\Bigg] , \label{2580c}
\\
&T(r)= -\frac{2}{b_0^2}\left[b_0^2\,r^{-2}+2\delta\,b_0\,(1+a)\,r^{-2-b}+(2\,a+1)\,r^{-2-2b}\right] . \label{2580d}
\\
&\tilde{a}_0\,r^a= \left[\left(1+\alpha\right)\rho^{1-w} +\beta\right]^{\left[\frac{\alpha}{\left(1+\alpha\right)\left(w-1\right)}-\frac{w}{w-1}\right]}\,\rho^{-w}  \label{2580f}
\end{align}
\end{subequations}
\normalsize
Because we are still looking for new $F(T)$ solutions, we need as in all previous cases to isolate $r(T)$ from eqn \eqref{2580d} leading to the following characteristic eqn:
\begin{align}\label{2580e}
0=\frac{b_0^2\,T}{2}+b_0^2\,r^{-2}+2\delta\,b_0\,(1+a)\,r^{-2-b}+(2\,a+1)\,r^{-2-2b}.
\end{align}
As in the section \ref{sect32}, there are three classes of solution according to $a$:
\begin{enumerate}
\item \textbf{General case (${\bf a \neq \left\lbrace-1,\,-\frac{1}{2}\right\rbrace }$)}: we can solve eqn \eqref{2580e} for $b=\left\lbrace 0,\,1,\,-1,\,-2 \right\rbrace $ leading to $r(T)$ solution in each case. Then we have to solve for each four value of $b$ the eqns \eqref{2580a} to \eqref{2580c} (the FEs) w.r.t the eqn \eqref{2252} (conservation laws). As a relevant case, we will solve the $b=0$ subcase for comparison with perfect and dust fluid solutions. First the eqn \eqref{2580f} for conservation law remains invariant and the eqn \eqref{2580e} will be:
\begin{align}\label{2589e}
& 0=\frac{b_0^2\,T}{2}\,r^2+\left(b_0^2+2\delta\,b_0\,(1+a)+2\,a+1\right),
\nonumber\\
&\Rightarrow r^2(T)=\frac{2\left(b_0^2+2\delta\,b_0\,(1+a)+2\,a+1\right)}{b_0^2\,(-T)},
\end{align}
where $T\leq 0$ and $\delta_1=\pm 1$. The eqns \eqref{2580a} to \eqref{2580c} will simplify as:
\small
\begin{subequations}
\begin{align}
& F'(-T)= -F'(0)\,\left[\frac{2\left(b_0^2+2\delta\,b_0\,(1+a)+2\,a+1\right)}{b_0^2}\right]^{\frac{\left(2a-a^2+1-b_0^2\right)}{2\left(a+1+\delta\,b_0\right)}}\,(-T)^{-\frac{\left(2a-a^2+1-b_0^2\right)}{2\left(a+1+\delta\,b_0\right)}} , \label{2589a}
\\
& \left(1+\alpha\right)\left(\kappa\rho\right)+\kappa^{1-w}\,\beta\,\left(\kappa\rho\right)^w = \frac{(-T)\,F'(T)}{\left(b_0^2+2\delta\,b_0\,(1+a)+2\,a+1\right)}\,\Bigg[-\frac{\left(2a-a^2+1-b_0^2\right)}{\left(a+1+\delta\,b_0\right)}\left(1+\delta\,b_0\right)+a\Bigg] , \label{2589b}
\\
&\kappa\,\rho = -\frac{F(T)}{2} + \frac{\left(1+\delta\,b_0\right)\,(-T)\,F'(T)}{\left(b_0^2+2\delta\,b_0\,(1+a)+2\,a+1\right)}\Bigg[-\frac{\left(2a-a^2+1-b_0^2\right)}{\left(a+1+\delta\,b_0\right)}-(a+1)\Bigg]. \label{2589c}
\end{align}
\end{subequations}
\normalsize
By substituting eqn \eqref{2589c} into eqn \eqref{2589b} and by expressing $F$ and $F'$ in terms of $(-T)$, we obtain a DE for pure $F(T)$ solution as a characteristic algebraic equation with $F'(-T)$ expressed by eqn \eqref{2589a}. This expression for $w=2$ and its solution will be expressed as:
\begin{align}
0=&G^2(F(-T),-T)+\frac{\kappa\left(1+\alpha\right)}{\beta}\,G(F(-T),-T)-C(-T),
\nonumber\\
&\Rightarrow\,G(F(-T),-T)=-\frac{\kappa\left(1+\alpha\right)}{2\beta} -\delta_1\sqrt{\left(\frac{\kappa\left(1+\alpha\right)}{2\beta}\right)^2+C(-T)} \label{2589g}
\end{align}
where $\delta_1=\pm 1$ and, 
\small
\begin{subequations}
\begin{align}
G(F(-T),-T)=& -\frac{F(-T)}{2} - \frac{\left(1+\delta\,b_0\right)F'(0)}{\left(b_0^2+2\delta\,b_0\,(1+a)+2\,a+1\right)}\Bigg[\frac{\left(2a-a^2+1-b_0^2\right)}{\left(a+1+\delta\,b_0\right)}+(a+1)\Bigg]
\nonumber\\
&\times\,\left[\frac{2\left(b_0^2+2\delta\,b_0\,(1+a)+2\,a+1\right)}{b_0^2}\right]^{\frac{\left(2a-a^2+1-b_0^2\right)}{2\left(a+1+\delta\,b_0\right)}}\,(-T)^{1-\frac{\left(2a-a^2+1-b_0^2\right)}{2\left(a+1+\delta\,b_0\right)}} , \label{2589ha}
\\
C(-T)=&-\frac{\kappa F'(0)}{\beta\left(b_0^2+2\delta\,b_0\,(1+a)+2\,a+1\right)}\,\Bigg[\frac{\left(2a-a^2+1-b_0^2\right)}{\left(a+1+\delta\,b_0\right)}\left(1+\delta\,b_0\right)-a\Bigg]
\nonumber\\
&\times\,\left[\frac{2\left(b_0^2+2\delta\,b_0\,(1+a)+2\,a+1\right)}{b_0^2}\right]^{\frac{\left(2a-a^2+1-b_0^2\right)}{2\left(a+1+\delta\,b_0\right)}}\,(-T)^{1-\frac{\left(2a-a^2+1-b_0^2\right)}{2\left(a+1+\delta\,b_0\right)}} \label{2589hb}
\end{align}
\end{subequations}
\normalsize
The eqn \eqref{2589g} solution will be for $F(-T)$ as:
\begin{align}\label{2589i}
F(-T) =& \frac{\kappa\left(1+\alpha\right)}{\beta} - \frac{2\left(1+\delta\,b_0\right)F'(0)}{\left(b_0^2+2\delta\,b_0\,(1+a)+2\,a+1\right)}\Bigg[\frac{\left(2a-a^2+1-b_0^2\right)}{\left(a+1+\delta\,b_0\right)}+(a+1)\Bigg]
\nonumber\\
&\times\,\left[\frac{2\left(b_0^2+2\delta\,b_0\,(1+a)+2\,a+1\right)}{b_0^2}\right]^{\frac{\left(2a-a^2+1-b_0^2\right)}{2\left(a+1+\delta\,b_0\right)}}\,(-T)^{1-\frac{\left(2a-a^2+1-b_0^2\right)}{2\left(a+1+\delta\,b_0\right)}} 
\nonumber\\
&+\delta_1 \Bigg[\left(\frac{\kappa\left(1+\alpha\right)}{\beta}\right)^2-\frac{4\kappa F'(0)}{\beta\left(b_0^2+2\delta\,b_0\,(1+a)+2\,a+1\right)}\nonumber\\
&\times\,\Bigg[\frac{\left(2a-a^2+1-b_0^2\right)}{\left(a+1+\delta\,b_0\right)}\left(1+\delta\,b_0\right)-a\Bigg]
\nonumber\\
&\times\,\left[\frac{2\left(b_0^2+2\delta\,b_0\,(1+a)+2\,a+1\right)}{b_0^2}\right]^{\frac{\left(2a-a^2+1-b_0^2\right)}{2\left(a+1+\delta\,b_0\right)}}\,(-T)^{1-\frac{\left(2a-a^2+1-b_0^2\right)}{2\left(a+1+\delta\,b_0\right)}}\Bigg]^{1/2} 
\end{align}
As expected, eqn \eqref{2589i} describes a complex non-linear perfect cosmological fluid teleparallel $F(T)$ solution. Here is a proof that non-linear fluids can lead to relevant solutions.

\item ${\bf a=b=0}$ special case: eqn \eqref{2580e} becomes:
\begin{align}\label{2581}
0=&\frac{b_0^2\,T}{2}+\left(1+\delta\,b_0\right)^2\,r^{-2},
\nonumber\\
&\Rightarrow\;r^{-2}(T)=-\frac{b_0^2\,T}{2\left(1+\delta\,b_0\right)^2}
\end{align}
By substituting eqn \eqref{2580c} into eqn \eqref{2580b} and then by putting eqn \eqref{2580a} inside, we find an algebraic equation:
\small
\begin{align}
0=&\Bigg(-\frac{F(T)}{2} + F'(0)\,\left[\frac{(-2)^{\frac{1-\delta\,b_0}{2}}\left(\alpha\,\left(2-\delta\,b_0\right)+1\right)}{\left(1+\alpha\right)b_0^{1-\delta\,b_0}\left(1+\delta\,b_0\right)^{\delta\,b_0}}\right]\,T^{\frac{\left(1+\delta\,b_0\right)}{2}}\Bigg)
\nonumber\\
&\;+\frac{\kappa^{1-w}\,\beta}{\left(1+\alpha\right)}\,\Bigg(-\frac{F(T)}{2} + F'(0)\,\left[\frac{(-2)^{\frac{1-\delta\,b_0}{2}}\left(2-\delta\,b_0\right)}{b_0^{1-\delta\,b_0}\left(1+\delta\,b_0\right)^{\delta\,b_0}}\right]\,T^{\frac{\left(1+\delta\,b_0\right)}{2}}\Bigg)^w , \label{2581d}
\end{align}
\normalsize
Eqn \eqref{2581d} is a degree $w$ polynomial equation in terms of $F(T)$ and it is solvable for $w=2,\,3$ and $4$. For $w=2$, eqn \eqref{2581d} is with simplifications:
\small
\begin{align}
0=&\Bigg(\frac{\beta}{2\kappa\left(1+\alpha\right)}\Bigg)\,F(T)^2+\Bigg(\frac{\beta\,F'(0)}{\kappa\left(1+\alpha\right)}\left[\frac{(-2)^{\frac{3-\delta\,b_0}{2}}\left(2-\delta\,b_0\right)}{b_0^{1-\delta\,b_0}\left(1+\delta\,b_0\right)^{\delta\,b_0}}\right]\,T^{\frac{\left(1+\delta\,b_0\right)}{2}}-1\Bigg)\,F(T) 
\nonumber\\
&\,+\Bigg(\frac{\beta\, F'(0)^2}{2\kappa\left(1+\alpha\right)}\left[\frac{(-2)^{\frac{3-\delta\,b_0}{2}}\left(2-\delta\,b_0\right)}{b_0^{1-\delta\,b_0}\left(1+\delta\,b_0\right)^{\delta\,b_0}}\right]^2\,T^{1+\delta\,b_0}- F'(0)\,\left[\frac{(-2)^{\frac{3-\delta\,b_0}{2}}\left(\alpha\,\left(2-\delta\,b_0\right)+1\right)}{\left(1+\alpha\right)b_0^{1-\delta\,b_0}\left(1+\delta\,b_0\right)^{\delta\,b_0}}\right]\,T^{\frac{\left(1+\delta\,b_0\right)}{2}}\Bigg) , \label{2581e}
\end{align}
\normalsize
The general solution is:
\begin{align}
F(T)=&\frac{\kappa\left(1+\alpha\right)}{\beta}\Bigg[1-\left(\frac{\beta\,F'(0)(-2)^{\frac{3-\delta\,b_0}{2}}\left(2-\delta\,b_0\right)}{\kappa\left(1+\alpha\right)b_0^{1-\delta\,b_0}\left(1+\delta\,b_0\right)^{\delta\,b_0}}\right)\,T^{\frac{\left(1+\delta\,b_0\right)}{2}}
\nonumber\\
&\,+\delta_2\sqrt{1+\left(\frac{\beta\,F'(0)(-2)^{\frac{5-\delta\,b_0}{2}}\left(1-\delta\,b_0\right)}{\kappa\left(1+\alpha\right)^2b_0^{1-\delta\,b_0}\left(1+\delta\,b_0\right)^{\delta\,b_0}}\right)\,T^{\frac{\left(1+\delta\,b_0\right)}{2}}}\Bigg]. \label{2581f}
\end{align}
Eqn \eqref{2581f} is a new teleparallel $F(T)$ solution and this function is not an approximated form. Hence for a flat teleparallel spacetime, there is a non-trivial $F(T)$ solution which is not a pure power-law solution.

\item ${\bf a=-1}$ case: eqn \eqref{2580e} will simplify as:
\begin{align}\label{2582}
0=\frac{b_0^2\,T}{2}+b_0^2\,r^{-2}-r^{-2-2b}.
\end{align}
Then we can solve this eqn \eqref{2582} for $b=\left\lbrace 0,\,\frac{1}{2},\,-\frac{1}{2},\,1,\,-1,\,-\frac{3}{2},\,2,\,-2,\,3 \right\rbrace$ as in section \ref{sect32}. After that, eqns \eqref{2580a} to \eqref{2580f} become for $a=-1$:
\small
\begin{subequations}
\begin{align}
F'(T)=& F'(0)\exp\left[-\frac{\delta}{b_0\,b}\left(-2\,r^{-b}(T)+b_0^2\,r^{b}(T)\right)\right]\quad\quad b\neq 0  , \label{2582aa}
\\
=& F'(0)\,\left[r(T)\right]^{-\frac{\delta\left(2+b_0^2\right)}{b_0}}\quad\quad\quad\quad\quad\quad\quad\quad\,\quad\quad\quad\, b= 0  , \label{2582ab}
\\
\left(1+\alpha\right)\,&\left(\kappa\rho\right)+\kappa^{1-w}\,\beta\,\left(\kappa\rho\right)^w = \frac{2}{b_0^2}\,F'(T)\,\Bigg[\frac{\left(2\,r^{-2b}+b_0^2\right)}{\delta\,b_0}\left(r^{-b}+\delta\,b_0\right)\,r^{-2}+(b-1)\,r^{-2b-2}\Bigg] , \label{2582b}
\\
\kappa\,\rho =& -\frac{F(T)}{2} + \frac{2}{b_0^2}\,F'(T)\,\Bigg[\frac{\left(2\,r^{-2b}+b_0^2\right)}{\delta\,b_0}\left(r^{-b}+\delta\,b_0\right)\,r^{-2}+b\,r^{-2b-2}\Bigg] , \label{2582c}
\\
\frac{\tilde{a}_0}{r}=& \left[\left(1+\alpha\right)\rho^{1-w} +\beta\right]^{\left[\frac{\alpha}{\left(1+\alpha\right)\left(w-1\right)}-\frac{w}{w-1}\right]}\,\rho^{-w}  \label{2582f}
\end{align}
\end{subequations}
\normalsize
We will only solve the eqns \eqref{2582ab} to \eqref{2582f} for $b=0$ subcase solution. Eqn \eqref{2582f} remains unchanged and eqn \eqref{2582} becomes:
\begin{align}\label{2595}
r^{-2}(T)=\frac{b_0^2}{2\left(1-b_0^2\right)}\,T.
\end{align}
By substituting eqn \eqref{2582c} and then eqn \eqref{2582ab} into eqn \eqref{2582b} for $b=0$ and $w=2$, we will obtain the algebraic equation on the eqn \eqref{2589g} form (we only change terms in $-T$ for $T$ terms !). The solution of this new eqn \eqref{2589g} will be:
\small
\begin{align}
F(T) =& \frac{2F'(0)\,\left(2+b_0^2\right)}{\delta\,b_0\left(1-\delta b_0\right)}\left[\frac{b_0^2}{2\left(1-b_0^2\right)}\right]^{\frac{\delta\left(2+b_0^2\right)}{2b_0}}\,T^{1+\frac{\delta\left(2+b_0^2\right)}{2b_0}}+\frac{\kappa\left(1+\alpha\right)}{\beta}
\nonumber\\
& +\delta_1\Bigg[\left(\frac{\kappa\left(1+\alpha\right)}{\beta}\right)^2- \frac{4\kappa\,F'(0)}{\beta}\,\left[\frac{b_0^2}{2\left(1-b_0^2\right)}\right]^{\frac{\delta\left(2+b_0^2\right)}{2b_0}}\Bigg[\frac{\left(2+b_0^2\right)}{\delta\,b_0\left(1-\delta\,b_0\right)}-\frac{1}{\left(1-b_0^2\right)}\Bigg]\,T^{1+\frac{\delta\left(2+b_0^2\right)}{2b_0}}\Bigg]^{1/2} \label{2595f}
\end{align}
\normalsize

\item ${\bf a=-\frac{1}{2}}$ case: eqn \eqref{2580e} will simplify as:
\begin{align}\label{2583}
0=\frac{b_0\,T}{2}+b_0\,r^{-2}+\delta\,r^{-2-b}.
\end{align}
We can solve this eqn \eqref{2583} for $b=\left\lbrace 0,\,1,\,-1,\,2,\,-2,\,-3,\,4,\,-4,\,6,\,-6,\,-8 \right\rbrace$ as in section \ref{sect32}. Then eqns \eqref{2580a} to \eqref{2580f} become for $a=-\frac{1}{2}$:
\small
\begin{subequations}
\begin{align}
& F'(T)= F'(0)\,\exp\left[\int\,dr\,\frac{\left[\left(-\frac{1}{4}+\frac{b}{2}\right)\,r^{-2b}-b_0^2\right]}{\left[\frac{1}{2}\,r^{1-2b}+\delta\,b_0\,r^{1-b}\right]}\right] , \label{2583a}
\\
&\left(1+\alpha\right)\,\left(\kappa\rho\right)+\kappa^{1-w}\,\beta\,\left(\kappa\rho\right)^w = \frac{2}{b_0^2}\,F'(T)\,\Bigg[\frac{\left[\left(\frac{1}{4}-\frac{b}{2}\right)\,r^{-2b}+b_0^2\right]}{r^2\left(\frac{r^{-b}}{2}\,+\delta\,b_0\right)}\left(r^{-b}+\delta\,b_0\right)+\left(b-\frac{1}{2}\right)\,r^{-2b-2}\Bigg] , \label{2583b}
\\
&\kappa\,\rho = -\frac{F(T)}{2} + \frac{2}{b_0^2}\,F'(T)\,\Bigg[\frac{\left[\left(\frac{1}{4}-\frac{b}{2}\right)\,r^{-2b}+b_0^2\right]}{r^2\left(\frac{r^{-b}}{2}+\delta\,b_0\right)}\left(r^{-b}+\delta\,b_0\right)+\left(b-\frac{1}{2}\right)\,r^{-2b-2}-\frac{\delta\,b_0}{2}\,r^{-b-2}\Bigg] , \label{2583c}
\\
&\frac{\tilde{a}_0}{\sqrt{r}}= \left[\left(1+\alpha\right)\rho^{1-w} +\beta\right]^{\left[\frac{\alpha}{\left(1+\alpha\right)\left(w-1\right)}-\frac{w}{w-1}\right]}\,\rho^{-w}  \label{2583f}
\end{align}
\end{subequations}
\normalsize
We will again solve the eqns \eqref{2583a} to \eqref{2583f} only for $b=0$ subcase solution. Eqn \eqref{2583f} remains unchanged and eqn \eqref{2583} becomes:
\begin{align}\label{2597}
r^{-2}(T)=\frac{\delta\,b_0}{2\left(1+\delta\,b_0\right)}\,(-T).  
\end{align}
By substituting eqn \eqref{2583c} and then eqn \eqref{2583a} into eqn \eqref{2583b} for $b=0$ and by setting $w=2$, we obtain another quadratic equation on the eqn \eqref{2589g} form. The solution of this relation will be:
\small
\begin{align}
F(T)=& \frac{2\delta\,F'(0)}{b_0}\Bigg[\frac{\left(\frac{1}{4}+b_0^2\right)}{\left(\frac{1}{2}+\delta\,b_0\right)}-\frac{1}{2}\Bigg]\left[\frac{\delta\,b_0}{2\left(1+\delta\,b_0\right)}\right]^{\frac{\left(\frac{1}{4}+b_0^2\right)}{\left(1+2\delta\,b_0\right)}}\,(-T)^{1+\frac{\left(\frac{1}{4}+b_0^2\right)}{\left(1+2\delta\,b_0\right)}}+\frac{\kappa\left(1+\alpha\right)}{\beta}
\nonumber\\
& -\delta_1\Bigg[\left(\frac{\kappa\left(1+\alpha\right)}{\beta}\right)^2+\frac{4\kappa\,\delta\,F'(0)}{\beta\,b_0}\Bigg[\frac{\left(\frac{1}{4}+b_0^2\right)}{\left(\frac{1}{2}+\delta\,b_0\right)}-\frac{1}{2\left(1+\delta\,b_0\right)}\Bigg]\left[\frac{\delta\,b_0}{2\left(1+\delta\,b_0\right)}\right]^{\frac{\left(\frac{1}{4}+b_0^2\right)}{\left(1+2\delta\,b_0\right)}}
\nonumber\\
&\,\times\,(-T)^{1+\frac{\left(\frac{1}{4}+b_0^2\right)}{\left(1+2\delta\,b_0\right)}}\Bigg]^{1/2} \label{2597h}
\end{align}
\normalsize

\end{enumerate}

All these previous $b=0$ teleparallel $F(T)$ solutions are expressing some possible cosmological spacetimes geometries. In the recent literature, there are some simple pure power-law $F(T)\sim (-T)^{k}$, logarithmic $F(T)\sim \ln (-T)$ or $F(T)\sim (-T)^{k}\,\ln (-T)$ leading to some stable solutions \cite{TdSpaper,Bahamonde:2021gfp,attractor}. In addition, we can do the same type of development for all $b \neq 0$ subcases and we may find some more complex non-linear fluid $F(T)$ solutions. As for sections \ref{sect3} and \ref{sect33}, some of these new solutions may lead to some black hole and/or matter absorbing singularity solutions by the end of this process. But in this current section, all necessary FEs and conservation law are there for further investigation in this way.


\section{Discussion and conclusion}

In this paper, we first solved conservation law and FEs and then found in section \ref{sect3}, \ref{sect33} and \ref{sect4} dozens of new teleparallel $F(T)$ solutions in static spherically symmetric spacetimes for perfect fluids. These new $F(T)$ solutions are products of exponential, power, quotients and some mix of these types of expression. In some of these new $F(T)$ solutions, we found some new singularities which arises to point-like discontinuity or undefined $F(T)$ functions. In section \ref{sect31}, we found new teleparallel $F(T)$ solutions for constant $A_3$ where we used a power-law ansatz in subsection \ref{sect311} and a special ansatz defined by a $A_2=$ constant (i.e. $b=0$ setting as in ref \cite{SSpaper}) and an exponential $A_1$ component in subsection \ref{sect312}. This $A_1$ component generalizes the power-law ansatz by a summation of an infinite number of integer power-law terms. By this approach, we found the same singularities as in subsection \ref{sect311} and an additional singularity arising from the new ansatz.

For the rest of the paper (subsections \ref{sect32} to \ref{sect43}), we used a power-law ansatz approach to find new $F(T)$ solutions by choosing a $A_3=r$ coordinate system. If $A_3\equiv $ constant, then we found for slightly quadratic perfect fluid approximation ($\beta \ll \alpha$ and $w=2$) some new approximated $F(T)$ solutions as showed by eqns \eqref{2571i} and \eqref{2571j} in section \ref{sect42}. The solutions found in section \ref{sect43} for non-linear perfect fluids (in particular $w=2$) are usually generalizing the power-laws $F(T)$ found in subsections \ref{sect322} to \ref{sect325} for $b=0$ and are exact. We can easily make the same assumptions for $b\neq 0$ cases for generalizing section \ref{sect32} new solutions. In addition, the new $F(T)$ solutions in section \ref{sect33} for cosmological dust fluids ($\alpha=0$) should be useful for studying some cosmological models with baryonic matter \cite{coleybook}.

Then we look for non-perfect fluids $F(T)$ solutions, but we will have at least to add supplementary terms to the eqn \eqref{1001d} definition of energy-momentum. We will have to add some factors such as viscosity and any fluid imperfections to only name them. Eqn \eqref{1001d} characterizes an ideal fluid without any viscosity or imperfection where the pressure and the density are directly linked by an EoS. But this assumption of eqn \eqref{1001d} cannot necessary be done for non-perfect fluids because these additional physical factors. Several works may be done in the future, but we can expect more complex $F(T)$ solutions than those found in this paper.

For astrophysical and cosmological applications, a detailed analysis for each $F(T)$ solutions obtained will be necessary for determining the stability conditions and their physical processes. There are several recent works on this type of study (See \cite{coleybook,attractor,paliathanasis2022f,Kofinas,leonpalia2,bahabohmer,ruggiero1,ruggiero2,coleylandrygholami} and references within). They sometimes replace the fluid by a scalar field source in some of these studies \cite{paliathanasis2022f,leonpalia2}. In addition, we should also study the physical processes around the singularities for each $F(T)$ solutions in some future works. We can also work with electromagnetic energy-momentum sources for new classes of $F(T)$ solutions and for possible ``electromagnetic'' BH horizons, but new $F(T)$ solutions will be necessary \cite{awad1,nashed5,pfeifer2,elhanafy1,benedictis3}. The teleparallel $F(T)$ solutions obtained in this paper can also be used as conditions for dynamical cosmological models. These solutions can be used for $(r,t)$-coordinates based $F(T)$ solutions in some astrophysical and cosmological problems. In addition, there are in this paper a lot of teleparallel $F(T)$ solutions for solving these physical problems and there are necessary ingredients for a complete cosmological analysis.

For going further on this approach, there are some ongoing developments concerning Kantowski-Sachs spacetimes solutions in teleparallel $F(T)$ gravity where we look for general, fluids and other solutions (See \cite{SSpaper} and references within). There are some possible works on axially symmetric teleparallel $F(T)$ geometries allowing to solve more astrophysical problems with teleparallel gravity \cite{bahamonde2021exploring,salvatore1}. Another possible works is by looking for teleparallel $F(T,B)$-type geometries. All these possibilities deserve serious and tactful considerations.

\section*{Acknowledgements}

AL is supported by an Atlantic Association of Research in Mathematical Sciences (AARMS) fellowship. Thanks to A. A. Coley, G. Leon and E. Gonzalez for discussions and comments. A special thanks to R. J. van den Hoogen for discussions and his very constructive and detailed comments.


\appendix

\section{Field equation components}\label{appena}

This appendix is for presenting the exact FE components found in ref \cite{SSpaper}. There are general, constant $A_3$ and $A_3=r$ power-law ansatz FE components for the current paper purposes. 

\subsection{General components}

\small
\begin{subequations}
\begin{align}
\frac{g_1}{k_1}=&\frac{\Bigg[-A_2\,A_3^2\,A_1''-A_1\,A_2\,A_3\,A_3''+A_1\,A_2\,A_3'^2+\left(A_1\,A_2\right)'\,A_3\,A_3'+A_3^2\,A_1'\,A_2'-A_1\,A_2^3\Bigg]}{\left[A_1\,A_2\,A_3\,A_3'+A_2\,A_3^2\,A_1'+\delta\,A_1\,A_2^2\,A_3\right]} \label{2922a}
\\
g_2 =& \frac{1}{A_1\,A_2^3\,A_3}\Bigg[-A_1\,A_2\,A_3''+\left(A_1\,A_2\right)'\,A_3'\Bigg] \label{2922b}
\\
g_3 =& \frac{1}{A_1\,A_2^3\,A_3^2}\Bigg[-A_1\,A_2\,A_3\,A_3''-A_1\,A_2\,A_3'^2-A_2\,A_3\,A_3'\,A_1'-\delta\,A_1\,A_2^2\,A_3'+A_1\,A_3\,A_2'\,A_3'-\delta\,A_2^2\,A_3\,A_1'\Bigg] ,\label{2922c}
\\
k_2=& k_3 = \frac{1}{A_2^2\,A_3}\left[A_3'+\delta\,A_2\right]  \label{2922e}
\end{align}
\end{subequations}
\normalsize

\subsection{$A_3=c_0=$ constant power-law components}

The eqns \eqref{2922a} to \eqref{2922e} with eqns \eqref{2160a} power-laws ansatz are:
\small
\begin{align}\label{2132i}
&\frac{g_1}{k_1} = \frac{\left[\left(a(1-a+b)\right)\,r^{-2b-2}-\left(\frac{b_0}{c_0}\right)^2\right]}{\left[a\,r^{-2b-1}+\delta\,\left(\frac{b_0}{c_0}\right)\,r^{-b}\right]} ,\quad  g_2=0 , \quad g_3 = -\left(\frac{\delta\,a}{b_0\,c_0}\right)\,r^{-b-1} , 
\quad k_2=k_3=\left(\frac{\delta}{b_0\,c_0}\right)\,r^{-b}  .
\end{align}
\normalsize

\subsection{$A_3=r$ power-law components}

The eqns \eqref{2922a} to \eqref{2922e} with eqns \eqref{2160a} power-laws ansatz are:
\small
\begin{align}\label{2923i}
&\frac{g_1}{k_1} = \frac{\left[\left(2a-a^2+ab+b+1\right)\,r^{-2b}-b_0^2\right]}{\left[(a+1)\,r^{2(1-b)-1}+\delta\,b_0\,r^{(1-b)}\right]} ,~
& g_2 = \frac{(a+b)}{b_0^2}\,r^{-2b-2} , &
\nonumber\\
& g_3 = \frac{1}{b_0^2}\Bigg[(-a+b-1)\,r^{-2b-2}-\delta\,b_0\,(a+1)\,r^{-b-2}\Bigg] ,~ & k_2= k_3 = \frac{1}{b_0^2}\Bigg[r^{-2b-1}+\delta\,b_0\,r^{-b-1}\Bigg] . &
\end{align}
\normalsize

\end{document}